\documentclass[12pt]{article}

\usepackage{newtxtext,newtxmath}

\usepackage{graphicx}

\usepackage[letterpaper,margin=1in]{geometry}

\renewenvironment{abstract}
	{\quotation}
	{\endquotation}

\date{}

\makeatletter
\renewcommand{\fnum@figure}{\textbf{Figure \thefigure}}
\renewcommand{\fnum@table}{\textbf{Table \thetable}}
\makeatother

\usepackage{scicite}

\usepackage{url}

\newcommand{\ket}[1]{\vert #1 \rangle}

\newcommand{\bra}[1]{\langle #1 \vert}

\newcommand{\braket}[2]{\langle #1 \vert #2 \rangle}

\newcommand{\ketHF}{\vert\text{HF}\rangle}
\newcommand{\braHF}{\langle\text{HF}\vert}
\newcommand{\ketT}[1]{\left|\text{#1}\right\rangle}

\renewcommand{\theta}{\vartheta}

\usepackage{bm}

\def\scititle{
	Resolving the ground state intersection problem in coupled cluster theory
}
\title{\bfseries \boldmath \scititle}

\author{
	Leo Stoll$^{1\dagger}$,
    Federico Rossi$^{1\dagger}$ and
	Henrik Koch$^{1\ast}$ \and
	\small$^{1}$Department of Chemistry, Norwegian University of Science and Technology, Trondheim 7050, Norway.\and
	\small$^\ast$Corresponding author. Email: henrik.koch@ntnu.no\and
	\small$^\dagger$These authors contributed equally to this work.
}

\begin{document} 

\maketitle

\begin{abstract} \bfseries \boldmath

A physically correct description of ground state conical intersections is essential for simulating non-adiabatic dynamics of radiationless decay to the electronic ground state. Coupled cluster theory offers a particularly balanced treatment of static and dynamic correlation in excited states. However, coupled cluster methods and other single-reference approaches fail to describe ground state intersection regions, and several workarounds have been proposed to address this shortcoming. Here we present convex similarity constrained coupled cluster theory (CVX-SCC), a framework that resolves the ground state intersection problem by construction. We have applied this theory to two different coupled cluster methods, and tested the resulting models on ethylene, uracil, PSB3, and HeH$_2$, showing that both variants produce physically correct ground state conical intersections. With future development of nuclear gradients and non-adiabatic coupling vectors, this approach will provide single-reference electronic structures suitable for non-adiabatic dynamics simulations all the way to the ground state.
\end{abstract}

\section*{Introduction}
The dynamical processes that occur after photoexcitation are of great interest in chemistry, biology and physics alike. Photoreactivity is exploited in biological processes as well as in chemical engineering. On the other hand, photostability is crucial to the longevity of desired chemical species. The central role light plays in the chemical processes that surround us has motivated both experimental and theoretical inquiries into the field of photodynamics. Radiative decay happens on a nanosecond time-scale, and has been studied extensively. In contrast, non-radiative relaxation processes are more intricate and much faster, occurring at the femtosecond time-scale. Among many, two interesting examples of relaxation by ultrafast internal conversion are non-radiative relaxation of nucleobases after absorption of ultraviolet radiation \cite{improta2016quantum}, and the biological pathways enabling vision in the human eye \cite{polli2015tracking}. 

During internal conversion, the nuclear wave-packet passes through degeneracies of electronic states, so called conical intersections, which allow varying degrees of non-adiabatic transfer to lower lying electronic states. This also implies that the nuclear and electronic motions become coupled. Although the existence of conical intersections has been known for almost a century \cite{TheCrossingOfPotentialSurfaces}, their ubiquitous role in photophysical and photochemical processes has only been discovered in recent decades \cite{NonadiabaticEventsandConicalIntersections, NonadiabaticityTheImportanceOfConicalIntersections, atchity1991potential, bernardi1990mechanism}. With the development of advanced pump-probe spectroscopy \cite{FemtosecondTimeResolvedPhotoelectronSpectroscopyofPolyatomicMolecules,  zewail2000femtochemistry}, today photodynamics can be studied experimentally at the femtosecond timescale in a tabletop lab setting \cite{smith2020femtosecond, worner2025ultrafast}. However, theoretical modeling of non-adiabatic dynamical processes poses a true challenge for theoretical chemists. Not only is a quantum treatment of the nuclear motion \cite{curchod2018ab, MolecularPhotochemistryRecentDevelopmentsinTheory} required, but the static description of the conical intersections themselves can be a challenge for standard electronic structure models. Different electronic-structure approaches can yield substantially different descriptions of conical intersections, motivating the continued development of reliable theoretical methods \cite{matsika2021electronic}. Coupled cluster theory is a highly attractive prospect, due to its detailed description of dynamical electron correlation.

Among conical intersections, degeneracies between the ground and first excited states are notoriously difficult to describe with single-reference methods. Near these ground state intersections the electronic ground state acquires an inherent multiconfigurational character, which single-reference methods are not built to capture. Additionally, the intermediate normalization commonly applied in single-reference methods prevents them from correctly describing the geometric phase effect \cite{berry1984quantal,GPECC} when circumnavigating such intersections. These failures manifest in several ways, including negative or complex excitation energies, presence of multiple solutions, artificial conical intersections, and discontinuous potential energy surfaces (PES). Affected methods include time-dependent Hartree-Fock (TDHF) \cite{MEST}, time-dependent density functional theory (TDDFT) \cite{casida1995time}, algebraic diagrammatic construction (ADC) \cite{PhysRevA.26.2395, ABTrofimov_1995, https://doi.org/10.1002/wcms.1206} methods and equation-of-motion coupled cluster (EOM-CC) \cite{EOM-CC} theory \cite{GPECC, tuna2015assessment}. The consequence is that highly correlated single-reference methods applied in excited state dynamics simulations, like second-order algebraic diagrammatic construction (ADC(2)) \cite{PhysRevA.26.2395, ABTrofimov_1995, https://doi.org/10.1002/wcms.1206} and the more recent similarity constrained coupled cluster singles and doubles (SCCSD) \cite{SCC, SCCSD, doi:10.1021/acs.jctc.4c00276, kjonstad2024photoinduced} models, fail when the dynamics approach the ground state.

The traditional solution to the ground state intersection problem is to employ multireference theories, like complete active space self-consistent field (CASSCF) \cite{ROOS1980157} or multireference configuration interaction (MRCI) \cite{doi:https://doi.org/10.1002/9781119417774.ch9} methods. However, these require careful selection of the active space. Moreover, due to computational limitations to the active-space size, they may not be feasible even for medium-sized systems. Notably, even some multireference perturbation methods, such as single-state complete active space second-order perturbation theory (SS-CASPT2) \cite{BATTAGLIA2023135, andersson1992second}, can incorrectly describe the topology of conical intersections when they do not treat the ground and excited states on equal footing \cite{gozem2014shape,park2019single}.

Given these limitations, several workarounds for single-reference models have been proposed \cite{matsika2021electronic}. One approach is to employ spin-flip formulations, in which both the ground and excited states are described as spin-flip excitations from a high-spin reference \cite{Krylov2001,Shao2003}. This approach allows the description of the multi-configurational character that arises near ground state conical intersections. The principal drawback is that these methods lead to spin-contamination, but recently mixed-reference spin-flip methods have been developed to solve this issue \cite{lee2018eliminating}. Although some additional limitations have been pointed out \cite{janoš2026limitations}, spin-flip methods remain a promising strategy for describing ground state intersections.

In this paper, we address the ground state intersection problem in the context of coupled cluster theory. Generalized coupled cluster (GCC) theory\cite{GCC} allows the description of the geometric phase effect and guarantees a unique solution that avoids discontinuities close to ground state intersections. Similarity constrained coupled cluster (SCC) theory \cite{SCC, SCCSD, SCC2} corrects unphysical artifacts caused by the non-Hermitian coupled cluster effective Hamiltonian \cite{StructureOptimizationForExcitedStatesWithCorrelatedSecondORderMethodsCC2ADC2, CrossingConditionsInCoupledClusterTheory, NonadiabaticityTheImportanceOfConicalIntersections, Cancoupledclustertheorytreatconicalintersections}, including complex energies and unphysical intersection topology and topography. However, SCC theory in its original form does not allow the use of the ground state as a similarity constrained state, and we therefore propose an alternative fixed-t SCC parameterization of the theory. By combining GCC and fixed-t SCC theory, we derive convex similarity constrained coupled cluster (CVX-SCC) theory and apply it at the coupled cluster singles and doubles (CCSD) \cite{CCSD} and the second-order approximate coupled cluster singles and doubles (CC2) \cite{CC2} levels of approximation. We demonstrate that the CVX-SCCSD and CVX-SCC2 models produce topologically correct ground state intersections, with minimal changes to the CCSD and CC2 energies. Importantly, convex similarity constrained methods maintain the computational scaling of the underlying standard coupled cluster models. While CVX-SCCSD includes the attractive CCSD treatment of dynamical electron correlation, the lower computational scaling of CVX-SCC2 will allow the study of larger molecular systems in non-adiabatic dynamics simulations to the ground state.

\section*{Results}\label{sec:results}
We initially validate CVX-SCC theory on an S$_0$/S$_1$ ground state conical intersection in ethylene. Then, we study an excited state conical intersection in uracil with the proposed fixed-t SCC models, and subsequently consider the viability of the CVX-SCCSD model to describe relaxation to the ground state. We move on to describing a ground state intersection in PSB3 using the CVX-SCC2 model, and compare its energy profiles to those of CC2 in a circular path around this intersection. Finally, we explore the use of CVX-HF as the reference for CVX-SCCSD near a ground state intersection in HeH$_2$. All results have been obtained from a modified development branch of the electronic structure program eT \cite{eT}.

\subsection*{Ethylene S$_0$/S$_1$ ground state conical intersection}

\begin{figure}[tb!]
    \centering
    \includegraphics[width=\linewidth]{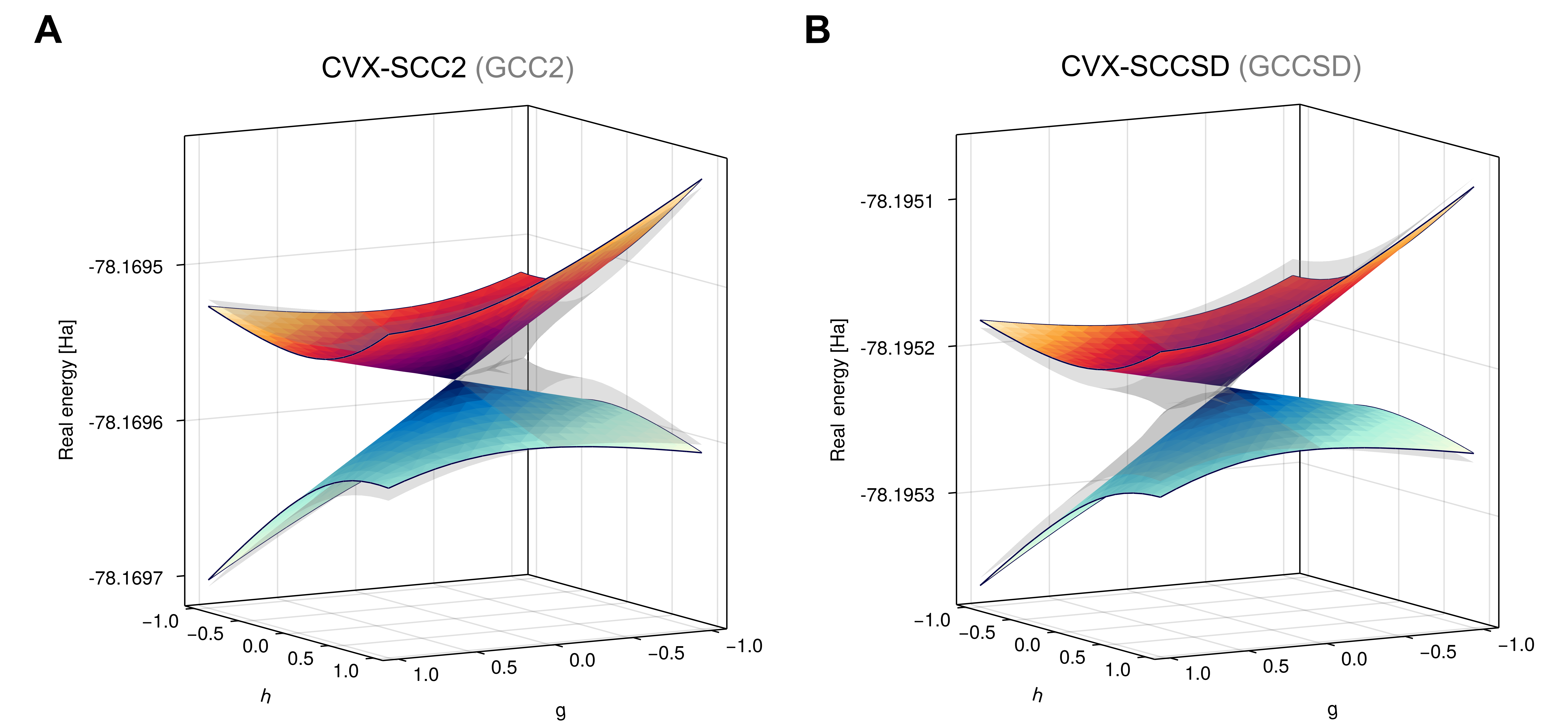}
    
    \caption{\textbf{Ethylene S$_0$/S$_1$ ground state conical intersections obtained with CVX-SCC theory (colored) and GCC theory (gray), using the aug-cc-pVDZ basis set.} Equations were converged to a residual threshold of $10^{-6}$ a.u.. The GCC conical intersections (gray) form a ring, corresponding to an $N-1$ dimensional intersection tube within the $N$ dimensional internal coordinate space. Within the intersection oval the electronic energies form a complex pair with degenerate real components. The CVX-SCC intersections (colored) display as a point, corresponding to an $N-2$ dimensional intersection seam, and no complex electronic energies are encountered. The coloring of the PES indicates the relative energy difference between the states. Note, that the coordinate systems are not the same in the two 2D-scans. The gh-plane was determined at the CCSD level \cite{angelico2025determining, kjonstad2023communication}. Details on the intersection geometries and g- and h-vectors are provided in the SI.}
    \label{fig:Ethylene_S0S1}
\end{figure}

The ethylene S$_0$/S$_1$ ground state intersection was previously used to test the GCCSD model \cite{GCC}. In Figure \ref{fig:Ethylene_S0S1}, we show 2D-scans of this intersection for CVX-SCC2 and CVX-SCCSD, together with their GCC counterparts. Due to the geometric phase effect, the standard coupled cluster models show severe unphysical artifacts in the intersection region. Most notably, these include negative excitation energies and discontinuities in the PES, and in some regions the CC calculations do not converge. Due to these issues, we omit the visualization of the standard CC ground state intersections. The CC2 and CCSD results are available in the data repository specified in the Data and materials availability section. 

The GCC models, on the other hand, have continuous PES with the correct ordering of the ground and excited states, as was already shown for GCCSD  \cite{GCC}. Our results confirm that the extension to GCC2 shows the same behavior. Still, the GCC models produce unphysical complex energy pairs within an intersection oval of wrong intersection dimensionality. Further, they display the typical non-linearity also seen in same-symmetry excited state intersections in standard coupled cluster theory. The CVX-SCC models rectify these flaws, producing ground state intersections with physically correct topology and topography. For both CVX-SCC2 and CVX-SCCSD the degeneracy is lifted linearly when moving away from the intersection point, avoiding complex energies in the region. In short, our calculations on ethylene demonstrate that CVX-SCC theory is able to produce qualitatively correct ground state intersections using a coupled cluster wave function parameterization.

\subsection*{Internal conversion in uracil}

\begin{figure}[tbp!]
    \centering
    \includegraphics[width=0.8\linewidth]{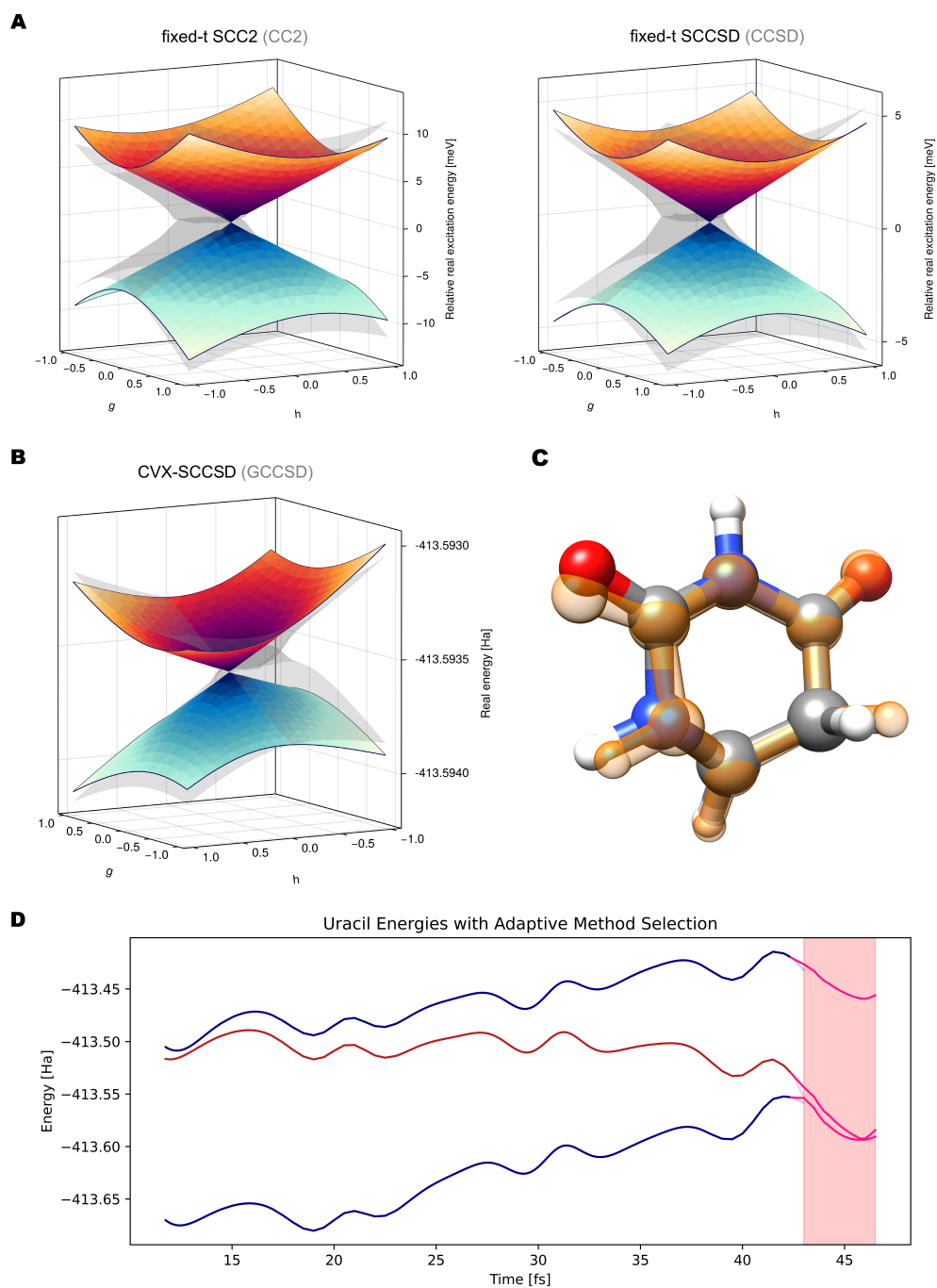}
    
    \caption{\footnotesize\setlength{\baselineskip}{8pt}\textbf{Uracil results with the cc-pVDZ basis set.} (A) S$_1$/S$_2$ conical intersections obtained from fixed-t SCC theory (colored) and standard CC theory (gray). Equations are converged to a residual threshold of $10^{-6}$ a.u.. Note that the gh-plane was determined at the CCSD level \cite{angelico2025determining, kjonstad2023communication} and the intersection point of each method was determined in this branching plane. The coordinate systems are not the same in the two 2D-scans. Details on the intersection geometries and applied g- and h-vectors are provided in the SI. (B) Uracil S$_0$/S$_1$ conical intersection of CVX-SCCSD theory (colored), overlayed with the GCCSD intersection (gray). Equations were converged to a residual threshold of $10^{-7}$ a.u.. The gh-plane was determined at the CCSD level \cite{angelico2025determining, kjonstad2023communication}. Details on the intersection geometries and g- and h-vectors are provided in the SI. (D) Uracil S$_0$, S$_1$ and S$_2$ PES of a trajectory obtained in a non-adiabatic dynamics simulation using \textit{ab initio} multiple spawning (AIMS) \cite{ben2000ab, ben2002ab, curchod2018ab} with electronic structures from an adaptive CCSD/SCCSD scheme \cite{kjonstad2024photoinduced, angelico2026spectroscopic}. The trajectory crashes around 43 fs into the simulation, as it approaches a ground state intersection where both CCSD and SCCSD break down. Red shading indicates geometries obtained by interpolating between the final geometry of the trajectory and the ground state intersection geometry. We stress that the time-coordinate has no physical meaning in the interpolated region. Red and blue graphs pertain to the CCSD/SCCSD energies, with the red graph indicating the active electronic state of the trajectory. Translucent graphs are used for the final geometries of the trajectory, where the CCSD results are considered unreliable. The pink graphs indicate CVX-SCCSD energies. Details on the calculations are provided in the SI. (C) Molecular geometries of the last timestep of the dynamics trajectory (orange) and S$_0$/S$_1$ conical intersection of CVX-SCCSD theory (colored).}
    \label{fig:Uracil}
\end{figure}

In uracil, after photoexcitation, ultrafast non-radiative relaxation to the ground state is facilitated by a high density of conical intersections in proximity of the ground state equilibrium geometry \cite{10.1021/jp048284n}. In Figure \ref{fig:Uracil} A, we consider an S$_1$/S$_2$ same-symmetry excited state intersection to validate the fixed-t SCC methods introduced in this work. The results for CC2 and CCSD, reported in gray, show a region of complex excitation energies within an intersection hypertube of wrong dimension and with a non-linear lifting of the degeneracy. This effect has previously been described \cite{CrossingConditionsInCoupledClusterTheory, SCC} and originates from the non-Hermitian coupled cluster effective Hamiltonian. In contrast, the potential energy surfaces of fixed-t SCC show the correct topology and linear lifting of the degeneracy, similar to the standard SCC models \cite{SCCSD, SCC2}. Our results indicate that fixed-t SCC theory is a viable alternative to SCC theory. 

Expanding the picture, we consider the internal conversion of uracil to the ground state, where a substantial part of the wave-packet ultimately passes through an S$_0$/S$_1$ same-symmetry ground state intersection. The difficulties with describing such intersections have complicated coupled cluster dynamics simulations to the ground state. In Figure \ref{fig:Uracil} D, we show the evolution of the energies for a trajectory obtained from an adaptive CCSD/SCCSD \cite{kjonstad2024photoinduced, angelico2026spectroscopic} non-adiabatic dynamics simulation using AIMS \cite{ben2002ab, ben2000ab, curchod2018ab}. Approximately 43 fs after photoexcitation, the dynamics simulation crashes as one of the trajectory basis functions approaches an S$_0$/S$_1$ ground state intersection, where both CCSD and SCCSD break down. We located the CVX-SCCSD S$_0$/S$_1$ conical intersection close to the last geometry of the dynamics simulation, which shows the correct topology as shown in Figure \ref{fig:Uracil} B. We then generated a series of geometries to interpolate between the last geometry of the dynamics and the geometry of the CVX-SCCSD intersection, including some points beyond the intersection. In the interpolated region (shaded red) in Figure \ref{fig:Uracil} D, the CVX-SCCSD results are continuous with positive excitation energies, and the intersection is described without issues. We point out that as energy gradients and non-adiabatic couplings have not yet been implemented for CVX-SCCSD, the results are not from a dynamics simulation but are used to demonstrate that CVX-SCC methods can be used to describe a trajectory that moves past an intersection with the ground state.

\subsection*{PSB3 S$_0$/S$_1$ ground state conical intersection}

\begin{figure}[tb!]
    \centering
    \includegraphics[width=\linewidth]{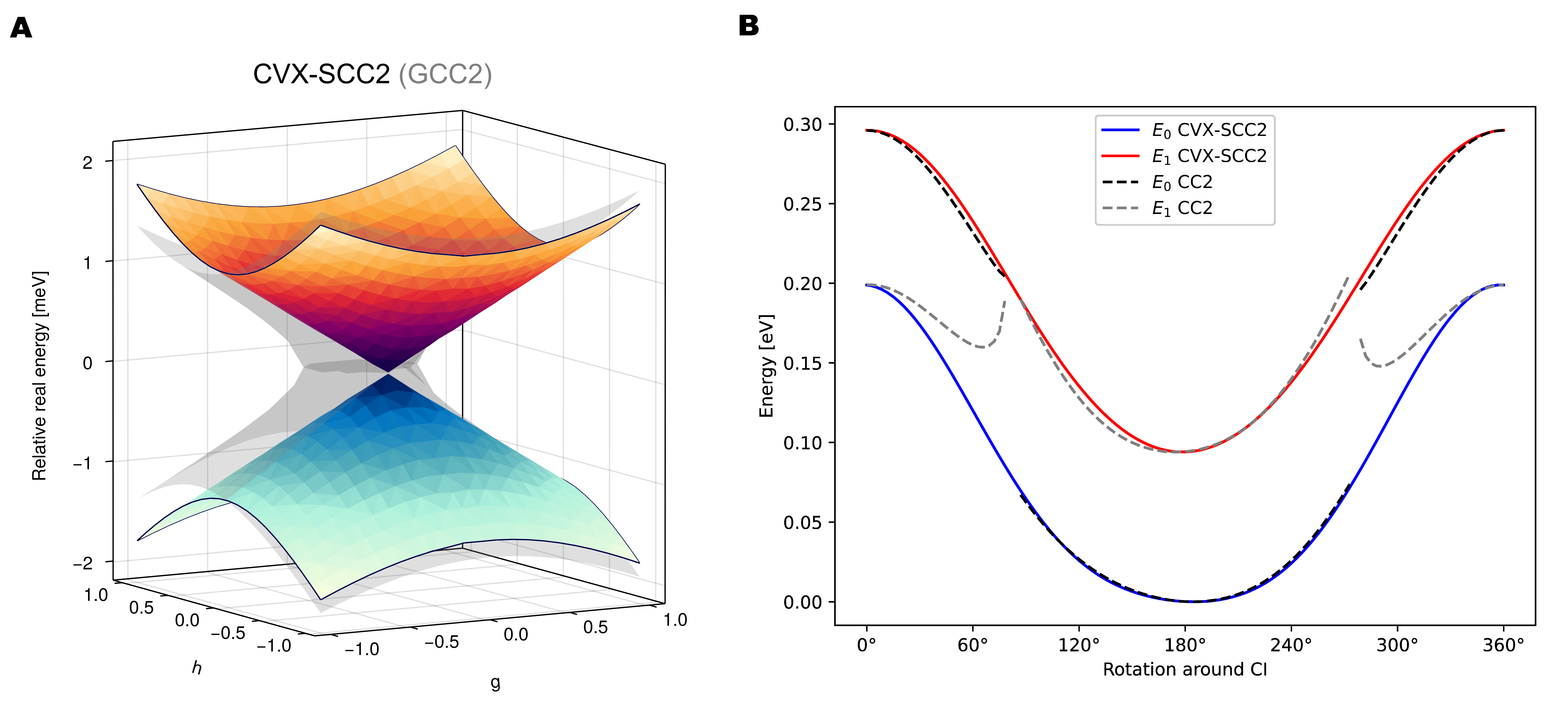}
    \caption{\textbf{PSB3 energy calculations with the 6-31G* basis set.} (A) S$_0$/S$_1$ ground state intersection of the CVX-SCC2 (colored) and GCC2 (gray) models. Equations were converged to a residual threshold of $10^{-6}$ a.u.. The gh-plane was determined at the CCSD level \cite{angelico2025determining, kjonstad2023communication}. Details on the intersection geometries and g- and h-vectors are provided in the SI. (B) S$_0$ and S$_1$ energy profiles of PSB3 when circumnavigating the CI. The CVX-SCC2 results are shown in blue ($E_0$) and red ($E_1$) and CC2 results in black ($E_0$) and gray ($E_1$). All energies are expressed in eV with respect to the minimum of the CVX-SCC2 energy. The circle is centered at the CVX-SCC S$_0$/S$_1$ intersection with a radius of 0.02\AA. A corresponding energy difference plot is shown in the SI.}
    \label{fig:PSB3_circle}
\end{figure}

Penta-2,4-dieniminium (PSB3) is a minimal model of retinal and has become a standard test system to investigate photoisomerization in vision-related photochemistry \cite{tuna2015assessment}. This motivates the desire to apply methods like ADC(2) or CC2 in non-adiabatic dynamics simulations of the molecule. As explored by Tuna et al. \cite{tuna2015assessment}, CC2 displays discontinuities and negative excitation energies in the intersection region. ADC(2) has similar issues, with negative excitation energies, and artificial ground state intersections. In Figure \ref{fig:PSB3_circle}, we show the PSB3 S$_0$/S$_1$ ground state intersection of CVX-SCC2, and compare the energy profiles to those of CC2 in a circular path around the intersection. In contrast to the unphysical CC2 and ADC(2) results in the literature \cite{tuna2015assessment}, CVX-SCC2 shows no discontinuities and recovers the correct state ordering in the circular path. 

\subsection*{HeH$_2$ S$_0$/S$_1$ ground state conical intersection}
\begin{figure}[tbp!]
    \centering
    \includegraphics[width=0.9\linewidth]{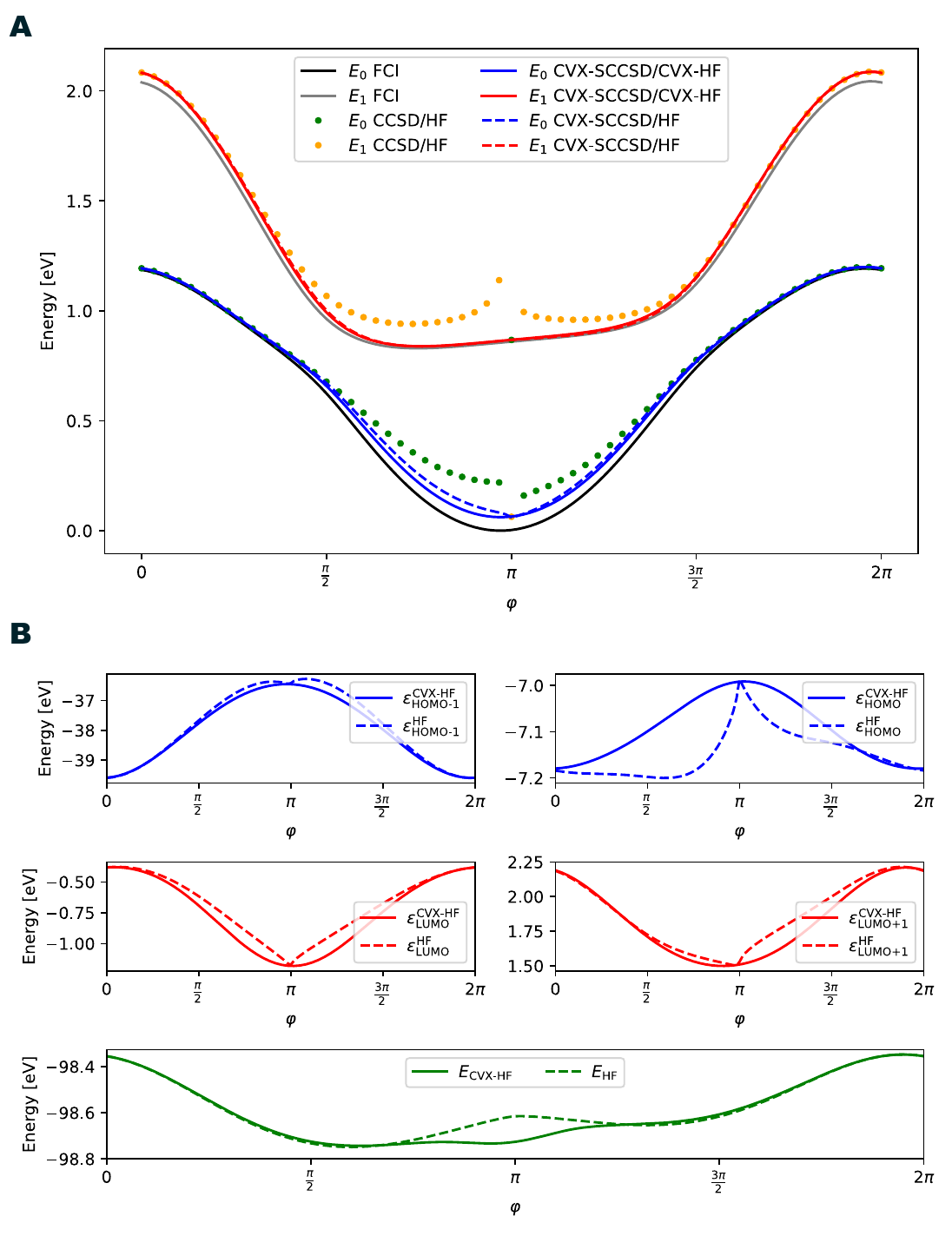}
    \caption{\setlength{\baselineskip}{8pt}\textbf{S$_0$ and S$_1$ energy profiles of HeH$_2$ when circumnavigating an S$_0$/S$_1$ conical intersection \cite{GPECC}, obtained with the aug-cc-pVDZ basis set.} (A) CVX-SCCSD results are shown in blue ($E_0$) and red ($E_1$), with full lines for CVX-HF and dashed lines for HF as the reference state. CCSD results are represented with green ($E_0$) and orange ($E_1$) dots and FCI with black ($E_0$) and gray ($E_1$) lines. All energies are expressed in eV with respect to the minimum of the FCI energy. Further details on the circular path and a 2D-scan of a corresponding CVX-SCCSD/CVX-HF ground state intersection are provided in the SI. (B) A selection of orbital energies are reported in full lines for CVX-HF and in dashed lines for HF, together with the total energies of the respective methods.}
    \label{fig:HeH2_circle}
\end{figure}

Until now, we have shown how the correct topology and physical topography of ground state conical intersections can be recovered using CVX-SCC methods. In the results reported above, Hartree-Fock was used as the reference determinant for all variants of coupled cluster calculations. We now consider the case of a ground state conical intersection in HeH$_2$, where Hartree-Fock has been reported to present cusps in the orbital energies \cite{GPECC}. In Figure \ref{fig:HeH2_circle} A, we compare the energy profiles of CCSD with FCI when moving around this intersection. The CCSD energies show artifacts due to the incorrect description of the phase effects, caused by intermediate normalization of the CC wave function. The use of CVX-SCCSD improves the description, but a small cusp in the energies is still present when HF is used as the reference. We believe that this is a leftover effect coming from the incorrect description at the Hartree-Fock level. In fact, the use of a CVX-HF reference restores full continuity of the solutions, eliminating all artifacts. The behavior of CVX-HF orbital energies can be seen in Figure \ref{fig:HeH2_circle} B, together with the HF results.

\section*{Discussion}

Convex similarity constrained coupled cluster theory combines GCC and SCC theories to handle challenges posed by the non-convexity of the amplitude equations and the non-Hermiticity of the coupled cluster effective Hamiltonian in one unified theory. For all the systems we considered, ethylene, uracil, PSB3, and HeH$_2$, the CVX-SCCSD and CVX-SCC2 models successfully describe ground state intersections. These results give us confidence that the model is robust enough for general applications. However, as illustrated for HeH$_2$, issues in the Hartree-Fock state may not be corrected by CVX-SCC theory. As such, care must be taken when selecting the reference state, and the use of the CVX-HF may be required in specific molecular systems.

In this analysis, we obtained conical intersection geometries within the standard CCSD branching planes. We note that a quantitative comparison requires branching planes at CVX-SCC level, which need access to CVX-SCC molecular gradients and non-adiabatic coupling vectors. These are not available at the moment, but can be obtained following the Lagrangian formalism applied in a previous work \cite{doi:10.1021/acs.jctc.4c00276} and will be the focus of future research.

The motivation for the development of CVX-SCC theory is its application in coupled cluster non-adiabatic dynamics simulations from excited states all the way to the ground state. In a previous application of SCC theory\cite{angelico2026spectroscopic}, an adaptive scheme has been applied for excited state non-adiabatic dynamics simulations using CCSD/SCCSD. This idea could be extended to include CVX-SCC theory, using standard CC as the baseline and switching to either fixed-t SCC or CVX-SCC when moving close to excited state or ground state conical intersections, respectively. This would allow dynamics simulations to run through any type of conical intersection at minimal computational cost. This type of adaptive scheme requires small variations in the electronic energies, molecular gradients and coupling vectors between the methods, to avoid sudden unphysical jumps in these values when switching models. In our experience, the differences in ground state and vertical excitation energies between CC, fixed-t SCC and CVX-SCC methods are small, and we expect this condition for an adaptive scheme to be met. Some examples can be found in Table \ref{tab:sup_EnergyComparison} in the SI.

\section*{Materials and methods}

\subsection*{Coupled cluster theory}

In coupled cluster theory, the ground state wave function $\ketT{CC}$ is given by the exponential ansatz
\begin{equation}
    \ketT{CC} = \exp(T) \ketHF
\end{equation}
acting on the reference state, usually the Hartree-Fock state $\ketT{HF}$. Here, $T$ is the coupled cluster operator
\begin{equation}
    T = \sum_{i=1}^{n_\text{cc}} T_i,
\end{equation}
expressed in terms of cluster operators $T_i$. Each one is a linear combination of excitation operators $\tau_{\mu_i}$ with rank $i$,
\begin{equation}
    T_i = \sum_{\mu_i} t_{\mu_i} \tau_{\mu_i}
\end{equation}
where $\tau_{\mu_i}$ generates the i-fold excited determinant $\ket{\mu_i} \equiv \tau_{\mu_i} \ketHF$ from the Hartree-Fock state, and $t_{\mu_i}$ is the corresponding cluster amplitude. The summation is over all excitations of rank $i$. The truncation level $n_\text{cc}$ determines the accuracy of the CC wave function. In the limit where $n_\text{cc}$ equals the number of electrons in the system, the CC wave function is identical to the full configuration interaction (FCI) wave function \cite{MEST}, with a few exceptions\cite{GPECC}. In practice, however, heavily truncated coupled cluster methods are applied. The most widely used methods are the coupled cluster singles and doubles (CCSD) and coupled cluster singles doubles and triples (CCSDT) \cite{CCSDT} methods, with truncation levels $n_\text{cc} = 2$ and $n_\text{cc} = 3$ respectively, and with respective computational cost scaling as $N^6$ and $N^8$ with the number of molecular orbitals $N$. Another popular method is the CC2 model, in which a perturbative treatment of the doubles equations is applied to obtain a computational scaling of $N^5$ \cite{CC2}.

The cluster amplitudes are determined from the amplitude equations
\begin{equation}
    \Omega_{\mu_i} \equiv\bra{\mu_i} \bar{H} \ketHF = 0
\end{equation}
for the projection manifold $\big\{\bra{\mu_i} \big| 0 < i \leq n_\text{CC}\big\}$. The $\bar{H} \equiv \exp(-T)H\exp(T)$ operator is the similarity transformed Hamiltonian, $H$ being the standard electronic Hamiltonian under the Born-Oppenheimer approximation \cite{MEST}. The coupled cluster ground state energy is determined from
\begin{equation}\label{eq:CC_energy}
    E_\text{CC} = \braHF \bar{H} \ketHF.
\end{equation}
From now on we will denote determinants in the projection manifold as $\bra{\mu}$ when the excitation order is not specified, and $\bra{\mu_0}$ will refer to the reference state.

Coupled cluster excitation energies are obtained from equation of motion coupled cluster theory (EOM-CC) \cite{EOM-CC}. The right and left excited states, $\ket{R^k}$ and $\bra{L^k}$ are expanded as linear combinations in the basis $\{ \ket{\mu_i} | 0 \leq i \leq n_\text{CC} \}$, that is
\begin{equation}
\begin{aligned}
    \ket{R^k} &= \sum_\nu \ket{\nu} r_{\nu}^k \\
    \bra{L^k} &= \sum_\mu \bra{\mu} l_{\mu}^k.
\end{aligned}
\end{equation}
The left and right excited state amplitudes, $l_{\mu}^k$ and $r_{\nu}^k$, are determined as the eigenvectors of the similarity transformed Hamiltonian matrix
\begin{equation}
    \mathbf{\bar{H}} = 
    \begin{pmatrix} 
        E_\text{CC} & \bm{\eta}^T \\
        \bm{0} & \mathbf{A} + E_\text{CC} \bm{I}
    \end{pmatrix},
\end{equation}
where $\mathbf{A}$ is the coupled cluster Jacobian \cite{MEST}
\begin{equation}
    A_{\mu\nu} = \bra{\mu} [\bar{H}, \tau_\nu] \ketHF
\end{equation}
and the $\bm{\eta}$ vector is
\begin{equation}
    \eta_\nu = \braHF [\bar{H}, \tau_\nu] \ketHF.
\end{equation}
Due to the block structure of $\mathbf{\bar{H}}$, the excitation energies $\omega_k \equiv E_k - E_0$ are obtained by solving the non-symmetric eigenvalue equations
\begin{equation}
\begin{aligned}
    \mathbf{A} \bm{r}^k &= \omega_k \bm{r}^k \\
    \mathbf{A}^T \bm{l}^k&= \omega_k \bm{l}^k,
\end{aligned}
\end{equation}
which exclude the ground state contributions $l^k_0$ and $r^k_0$. As the right and left states are neither orthonormal nor adjoint, calculations with these states are simplified by imposing biorthonormality ($\braket{L^k}{R^l} = \delta_{kl}$) explicitly. The ground state contributions to the excited states are determined as
\begin{equation}
\begin{aligned}
    r^k_0 &= \bm{\eta}^T\mathbf{A}^{-1}\bm{r}^k= -\bar{\bm{t}}^T\bm{r}^k\\
    l^k_0 &= 0,
\end{aligned}
\end{equation}
where $\bar{\bm{t}}$ is implicitly defined.

In spin-adapted singlet coupled cluster methods a biorthonormal basis for the bras $\{ \bra{\bar{\mu}}\}$ (satisfying $\braket{\bar{\mu}}{\nu} = \delta_{\mu\nu}$) is commonly applied \cite{MEST}. In the following we will let $\{ \bra{\mu}\}$ refer to the biorthonormal rather than the adjoint bra basis.

\subsubsection*{Convex Hartree-Fock theory}
Traditional mean-field frameworks like Hartree–Fock (HF) and density functional theory (DFT) and their linear-response time-dependent extensions struggle to describe ground state conical intersections. Convex Hartree-Fock (CVX-HF) theory \cite{CVXHF} was recently developed to overcome this problem. We start from the exponential parameterization of Hartree-Fock with a reference determinant $\ket{\Phi_0}$
\begin{align}
    \ket{\mathrm{HF}} = \exp \left(\sum_{ai} \kappa_{ai}E^-_{ai} \right) \ket{\Phi_0} ,\label{eq:Hartree-Fock}
\end{align}
where $\kappa_{ai}$ are the orbital rotation parameters of a single global orbital rotation. The projection operators $P_i = \ket{R^i}\bra{R^i}$ are constructed from the eigenvectors $\mathbf{r}_i$ of the electronic Hessian $\mathbf{G}^{(1)}$ with respect to an additional orbital rotation $\bm{\gamma}$
\begin{equation}
G^{(1)}_{ai,bj}(\mathbf{0}) = \frac{\partial^2 \mathcal{E}}{\partial \gamma_{ai}\partial \gamma_{bj}}\Big|_{\bm\gamma=\mathbf{0}}=\frac{1}{2}P_{ai,bj}\bra{\mathrm{HF}}[[H,E^-_{ai}],E^-_{bj}]\ket{\mathrm{HF}}, \label{eq:Hessian_HF}
\end{equation}
where $P_{ai,bj}$ permutes the indices such that $P_{ai,bj}A_{ai,bj}=A_{ai,bj}+A_{bj,ai}$.
These define a subspace 
\begin{align}
    \bra{\tilde{\mu}} &= \bra{\mu} - \sum_{i=1}^n \bra{\mu} P_i = \bra{\mu} - \sum_{i=1}^n r^i_\mu \bra{R^i},    \label{eq:projected_manifold_hf1}\\
    \ket{\tilde{\nu}} &= \ket{\nu} - \sum_{i=1}^n P_i \ket{\nu}  = \ket{\nu} - \sum_{i=1}^n  \ket{R^i}r^i_\nu
    \label{eq:projected_manifold_hf2}
\end{align}
and the corresponding set of coupled equations reads
\begin{align}
&\tilde{\mathbf{G}}^{(0)} \equiv \mathbf{G}^{(0)} - \sum_{i=1}^n P_i \mathbf{G}^{(0)}=0 ,\\
& \mathbf{G}^{(1)}\mathbf{r}^i = \lambda_i \mathbf{r}^i \hspace{15pt} i = 1,\dots,n
\end{align}
where $\mathbf{G}^{(0)} $ is the gradient of the energy with respect to an orbital rotation. The orbital rotation parameters are also projected $\tilde{\bm\kappa} = \bm\kappa -\sum_{i=1}^n P_i \bm\kappa$ and optimized, avoiding all issues due to degeneracies. The ground and excited state energies are retrieved by solving the eigenvalue problem for the Hamiltonian matrix in the full manifold
\begin{equation}
    \mathbf{H} = 
    \begin{pmatrix} 
        E_\text{HF}  & \mathbf{G}^{(0)\,T}_P \\
        \mathbf{G}^{(0)}_P & \mathbf{A} + E_\text{HF} \bm{I}
    \end{pmatrix},
\end{equation}
where the leftover contribution to the gradient is given by
\begin{equation}
\mathbf{G}^{(0)}_P = \sum_{i=1}^n P_i \mathbf{G}^{(0)}.
\end{equation}
Convex Hartree-Fock can be used as a reference for coupled cluster methods as long as the contribution of $\mathbf{G}^{(0)}_P$, usually missing because of Brillouin theorem, is explicitly included. Orbital energies and coefficients are obtained by diagonalizing the occupied-occupied and virtual-virtual blocks of the Fock matrix, i.e. setting the off-diagonal blocks to zero.

\subsection*{Generalized coupled cluster theory}
Standard coupled cluster theory produces unphysical bifurcations in the energies when moving around a ground state conical intersection, a consequence of the intermediate normalization of the wave function conflicting with the geometric phase\cite{longuet1958studies,berry1984quantal,GPEdiab}. This is accompanied by a region of negative eigenvalues for the coupled cluster Jacobian matrix\cite{GPECC}. Generalized coupled cluster theory (GCC)\cite{GCC} removes these artifacts and recovers continuous energies.

Consider the biorthonormal projection $P_i = \ket{R^i}\bra{L^i}$, constructed from the left and right eigenstates of the Jacobian matrix. When the CC Jacobian matrix is diagonalizable, then the projection operators $\{P_i\}$ form a complete set in the projection space, and we can write  
\begin{equation}
    I = \sum_i P_i.
\end{equation}
These projection operators can be used to selectively remove from the excitation manifold the components of all $n$ Jacobian eigenvectors associated with small or negative eigenvalues. This results in a subspace analogous to that of Eqs. \ref{eq:projected_manifold_hf1}-\ref{eq:projected_manifold_hf2},
\begin{align}
    \bra{\tilde{\mu}} &= \bra{\mu} - \sum_{i=1}^n \bra{\mu} P_i = \bra{\mu} - \sum_{i=1}^n r^i_\mu \bra{L^i},    \label{eq:projected_manifold1}\\
    \ket{\tilde{\nu}} &= \ket{\nu} - \sum_{i=1}^n P_i \ket{\nu}  = \ket{\nu} - \sum_{i=1}^n  \ket{R^i}l^i_\nu.
    \label{eq:projected_manifold2}
\end{align}
In this subspace, the Jacobian becomes positive definite, making the optimization of the amplitude equations a convex problem \cite{GCC}. The cluster amplitudes $\ket{t} = \sum_{\mu} \ket{\mu}t_{\mu}$ are also projected as
\begin{equation}
    \ket{t} = \ket{t'} - \sum_{i=1}^n \ket{R^i}\braket{L^i}{t'},
\end{equation}
and are determined by self-consistently solving the set of coupled equations
\begin{align}\label{eq:GCC_t}
    &\tilde{\Omega}_{\mu} \equiv \bra{\tilde{\mu}} \bar{H} \ketHF = 0\\
    &\mathbf{A}\bm{r}^i = \omega_i\bm{r}^i \hspace{15pt} i = 1,\dots,n\\
    &\mathbf{A}^T\bm{l}^i = \omega_i\bm{l}^i \hspace{15pt} i = 1,\dots,n .
\end{align}
Since components of the excitation space have been removed, the FCI limit can no longer be reached. The limit can be obtained returning to the full manifold $\{ \ketHF, \ket{\mu} \}$, where the similarity transformed Hamiltonian matrix $\mathbf{\bar{H}}$ has the block structure
\begin{equation}
    \mathbf{\bar{H}} = 
    \begin{pmatrix} 
        E_\text{CC}  & \bm{\eta}^T \\
        \bm{\Omega}^{P} & \mathbf{A} + E_\text{CC} \bm{I} + \mathbf{W}
    \end{pmatrix}.
\end{equation}
The Hamiltonian contains contributions from the projected components, which result in a leftover term in $\bm{\Omega}$ 
\begin{equation}
    \Omega_{\mu}^{P} \equiv \sum_{i=1}^n r^i_\mu \bra{L^i}\bar{H}\ketHF, 
\end{equation}
and an additional term in the excited-excited block
\begin{equation}
    W_{\mu \nu} \equiv \bra{\mu}\tau_\nu \bar{H}\ketHF - E_\text{CC} \, \delta_{\mu \nu} \,.
\end{equation}
Solving the eigenvalue problem
\begin{equation}
    \mathbf{\bar{H}} \bm{x}^k = E_k \bm{x}^k, \hspace{15 pt} k = 0,1,\dots\,
\end{equation}
we obtain energies $E_k$ and states $\bm{x}^k$ that recover contributions from the complete excitation space. We emphasize that both the ground state and the excited states are updated through this final diagonalization step.

We consider the application of GCC theory at the CCSD and CC2 levels. The projection manifold is then spanned by single and double singlet excitations of the reference state. As the reference, we choose either the HF state, or the CVX-HF state if stated explicitly. The GCCSD model has already been described in detail\cite{GCC}, and the extension to the GCC2 model is straightforward and no new terms have to be implemented, truncating the usual terms to obtain the corresponding GCC2 expressions \cite{CC2} from the GCCSD equations. The computational cost of the GCCSD and GCC2 models scales as $N^6$ and $N^5$ respectively, but with a different prefactor with respect to the standard CCSD and CC2 models.

\subsection*{Similarity constrained coupled cluster theory}
The similarity transformed Hamiltonian in coupled cluster theory is non-Hermitian. As a consequence, complex excitation energies can appear in the proximity of same-symmetry conical intersections. In such cases, the intersections themselves are determined by crossing conditions \cite{CrossingConditionsInCoupledClusterTheory} different from those of Hermitian theories \cite{TheCrossingOfPotentialSurfaces, UberDasVerhaltenVonEigenwertenBeiAdiabatischenProzessen}, and the intersections do not display the correct topology and topography. 

To restore the correct description, it is sufficient to require the intersecting states $\ket{R^k}$ and $\ket{R^l}$ (referred to as the similarity constrained states) to be orthogonal with respect to a positive-definite metric $M$ \cite{CrossingConditionsInCoupledClusterTheory, SCC},
\begin{equation}
    O(k,l)\equiv\bra{R^k}M\ket{R^l} = 0 \hspace{15 pt} \text{for} \hspace{15 pt} k\neq l,
\end{equation}
referred to as the orthogonality condition \cite{SCCSD}. This condition implies the diagonalizability of the coupled cluster Jacobian matrix in the subspace spanned by these states. Here we have assumed two similarity constrained states, though in principle it is possible to extend the theory to three or more intersecting states. To account for the additional constraint given by the orthogonality condition, a second similarity transformation with an excitation operator $X$, containing an additional wave function parameter $\zeta$, is applied and the SCC effective Hamiltonian is obtained
\begin{equation}\label{eq:SCC_H_tilde}
    \tilde{H} \equiv \exp(-X)\bar{H}\exp(X).
\end{equation}
The full set of coupled equations
\begin{equation}
\begin{aligned}
    \bm{\Omega}^S &= 0\\
    \mathbf{A}^S \bm{r}^k - \omega_k \bm{r}^k &= 0\\
    \mathbf{A}^S \bm{r}^l - \omega_l \bm{r}^l &= 0\\
    O(k,l) &= 0
\end{aligned}
\end{equation}
is solved self-consistently \cite{SCCSD}. Where we use the superscript $S$ to indicate the use of the SCC effective Hamiltonian. For example, the omega equations and the Jacobian are defined as ${\Omega}_{\mu}^S = \bra{\mu}\tilde{H}\ketHF$ and $\mathbf{A}_{\mu\nu}^S = \bra{\mu}[\tilde{H}, \tau_\nu]\ketHF$. Once the SCC equations are solved, the ground state energy is obtained in the usual way, as
\begin{equation}\label{eq:SCC_energy}
    E^S_\text{CC} = \braHF\tilde{H}\ketHF.
\end{equation}
Similarity constrained coupled cluster theory has been implemented for CCSD and CC2\cite{SCCSD, SCC2}, using the additional excitation operator
\begin{equation}
     X = X_3 = \zeta \sum_{\mu_1\mu_2} \left( r^k_{\mu_1} r^l_{\mu_2} - r^l_{\mu_1} r^k_{\mu_2} \right) \tau_{\mu_1} \tau_{\mu_2}.\label{eq:X3_SCC}
\end{equation}
With this choice of $X_3$ the inclusion of the ground state as a similarity constrained state is not possible. In the orthogonality condition, multiple metrics have been applied. In this work we focus on the natural projection
\begin{equation}
    M = \sum_\mu \exp(T)^\dagger\ket{\mu}\bra{\mu}\exp(T).\label{eq:M_natural} 
\end{equation}

\subsubsection*{Fixed-t similarity constrained coupled cluster theory}
To simplify the inclusion of the ground state as a similarity constrained state, we propose a variation of the original SCC theory presented above, in which the similarity transformation with $X$ is not applied to the amplitude equations. This ensures that the t-amplitudes are the same as in standard CC theory. States and energies are modified exclusively by diagonalization of the SCC effective Hamiltonian, which contains $X$, in the full projection manifold,
\begin{equation} \label{eq:SCC_H_ev_problem}
\begin{aligned}
    \mathbf{\tilde{H}} \bm{r}^k = E_k \bm{r}^k,
\end{aligned}
\end{equation} 
where $\tilde{H}_{\mu\nu} = \bra{\mu}\tilde{H}\ket{\nu}$. The SCC effective Hamiltonian $\mathbf{\tilde{H}}$ has the block structure
\begin{equation}
    \mathbf{\tilde{H}} = 
    \begin{pmatrix} 
        E^S_\text{CC}  & (\bm{\eta}^S)^T \\
        \bm{\Omega}^S & \mathbf{A}^S + E^S_\text{CC} \bm{I} + \mathbf{W}^S
    \end{pmatrix}.
\end{equation}
Note that, since we are working with the Hamiltonian matrix, coefficient vectors $\bm{r}^k$ and $\bm{l}^k$ here include the ground state contribution. By solving Eq. \ref{eq:SCC_H_ev_problem}, we obtain the ground state as 
\begin{equation}
    \ket{R^0} = \sum_{\mu} \ket{\mu} r_\mu^0,
\end{equation}
with an energy $E_0$. We stress that this energy is not equal to $E^S_\text{CC}$ determined by Eq. \ref{eq:SCC_energy}. If $X$ contains only triple or higher excitations $E^S_\text{CC} = E_\text{CC}$.

For two similarity constrained states $\ket{R^k}$ and $\ket{R^l}$, the set of coupled equations is
\begin{equation}\label{eq:SCC}
\begin{aligned}
    \mathbf{\tilde{H}} \bm{r}^k - E_k \bm{r}^k &= 0\\
    \mathbf{\tilde{H}} \bm{r}^l - E_l \bm{r}^l &= 0\\
    O(k,l) &= 0,
\end{aligned}
\end{equation}
where one of the similarity constrained states may be the SCC ground state ($k=0$). We would like to note that the fixed-t SCC parameterization will still produce ground state energies and excitation energies with correct size-scaling, as long as the additional excitations in $X$ are localized on one subsystem and the orthogonality condition itself has the correct size-scaling properties.

In this work, we consider the application of fixed-t SCC theory at the CCSD and CC2 levels of approximation. The projection manifold is then spanned by single and double singlet excitations of the reference state, which we here assume to be a HF determinant. To ensure orbital invariance and the FCI limit, we use the triple excitation operator and the natural projection metric of standard SCC as presented in Eq. \ref{eq:X3_SCC}-\ref{eq:M_natural}, respectively. All required terms have been derived and implemented previously, and we refer to the original publications of SCCSD and SCC2 for details \cite{SCCSD, SCC2}.

To solve the fixed-t SCC equations, we use a solver which combines a DIIS and Davidson solver. The DIIS solver is based on the solver outlined by Kjønstad and Koch \cite{SCCSD}, from which the amplitude equations are removed. After convergence of the DIIS solver, the Davidson solver \cite{davidson197514} is run with the optimized value for $\zeta$ and the DIIS vectors as the start guess. This step provides the modified SCC ground state, and additionally serves as a check for potential issues typical for DIIS solvers. The fixed-t SCC methods maintain the computational scaling of the original SCC methods, with a different prefactor. Specifically, fixed-t SCC2 and fixed-t SCCSD scale as $N^5$ and $N^6$ respectively.

\subsection*{Convex similarity constrained coupled cluster theory}

Near ground state intersections, generalized coupled cluster theory removes bifurcations and jumps in the PES by ensuring that the amplitude equations form a convex optimization problem. However, when looking at GCC ground state intersections in detail, an intersection hypertube of wrong dimensionality is observed. When moving away from the tube, the degeneracy is not lifted linearly, resulting in an intersection that is not actually conical. Finally, within the tube, the electronic energies become complex. Much like for same-symmetry excited state conical intersections in standard coupled cluster theory, these artifacts are caused by linear dependencies between the intersecting states, allowed by the non-Hermitian coupled cluster effective Hamiltonian. In SCC theory linear independence of the similarity constrained states is directly enforced, and in this way the topological and topographical characteristics of Hermitian theories are recovered. We can directly apply fixed-t SCC theory as a correction to the GCC results. We name the resulting model convex similarity constrained coupled cluster theory (CVX-SCC). This theory produces ground state intersections with physical topology and topography within a coupled cluster framework.

We start out from the GCC ground state amplitudes (Eq. \ref{eq:GCC_t}). The CVX-SCC effective Hamiltonian is then constructed in the following way. First, we apply the similarity transformation with $X$ directly to the GCC Hamiltonian. The matrix representation of the resulting Hamiltonian in the full manifold $\{ \ketHF, \ket{\mu}\}$ can be written as the sum of the original GCC Hamiltonian matrix $\mathbf{\bar{H}}$, and an SCC correction matrix $H^X_{\mu\nu} = \bra{\mu} \mathbf{\tilde{H}} - \mathbf{\bar{H}} \ket{\nu}$, that is
\begin{equation}
    \mathbf{\bar{H}} + \mathbf{H}^X=
    \begin{pmatrix} 
        E_\text{CC}  & \bm{\eta}^T \\
        \bm{\Omega}^P & \mathbf{A} + E_\text{CC} \mathbf{I} + \mathbf{W}
    \end{pmatrix} +
    \begin{pmatrix} 
        E_\text{CC}^X & (\bm{\eta}^\text{X})^T \\
        \bm{\Omega}^X& \mathbf{A}^X+ E_\text{CC}^X\mathbf{I} + \mathbf{W}^X
    \end{pmatrix}.
\end{equation}
The superscript $X$ is used to indicate that only the additional SCC contributions, containing $X$, are included in each term. We now apply a projection corresponding to that applied in GCC, acting on $\mathbf{H}^X$, to obtain the CVX-SCC Hamiltonian matrix
\begin{equation}
    \mathbf{\tilde{H}} = \mathbf{\bar{H}} + \sum_{i=1}^n \mathbf{P}_i \mathbf{H}^X \mathbf{P}_i,
\end{equation}
where
\begin{equation}
    \mathbf{P}_i = \begin{pmatrix}
        1 & 0 \\ 0 & P_i
    \end{pmatrix}.
\end{equation}
The projection avoids changes to the projected amplitude equations $\tilde{\bm{\Omega}}$ (see Eq. \ref{eq:GCC_t}), which ensures that the GCC t-amplitudes remain a solution of the projected amplitude equations with the CVX-SCC Hamiltonian. Finally, we solve the SCC set of equations (see Eq. \ref{eq:SCC}) with this effective Hamiltonian, determining $\zeta$ and the CVX-SCC states and energies.

We apply the convex similarity constrained framework to both CCSD (CVX-SCCSD) and CC2 (CVX-SCC2). As the reference, we choose either the HF state, or the CVX-HF state if stated explicitly. When selecting the additional excitation operator $X$, some additional challenges arise for ground state intersections. Only the singles and doubles components of the excited state coefficient vectors enter in $X_3$ for the similarity. The reference contribution may dominate in one of the similarity constrained states, leading to minimal singles and doubles contributions. Therefore, we employ the renormalized GCC full matrix states $\bm{x}^k_\mu$ (with the renormalization excluding the reference contribution) in $X_3$, which now reads
\begin{equation}
     X = X_3 = \zeta \sum_{\mu_1\mu_2} \left( x^k_{\mu_1} x^l_{\mu_2} + x^l_{\mu_1} x^k_{\mu_2} \right) \tau_{\mu_1} \tau_{\mu_2}.
\end{equation}
Note that to avoid cancellation of the two terms in geometries where the GCC full matrix states are parallel, we apply their plus combination. We stress that this choice of $X_3$ is totally symmetric for similarity constrained states of the same symmetry, and maintains orbital invariance and the correct size-scaling properties of the ground state energy and excitation energies. In the orthogonality condition, we apply the natural projection.

Furthermore, we approximate the GCC projection by using only the singles components of the EOM-CC states, that is
\begin{equation}
    \mathbf{\tilde{H}} = \mathbf{\bar{H}} + \sum_{i=1}^n P_{i}^{\mu_1} \,  \mathbf{H}^X\,P_i^{\mu_1},
\end{equation}
with $P_i^{\mu_1} = \ket{R^i_{\mu_1}}\bra{L^i_{\mu_1}}$. We find that the singles-only projection is both more robust and more efficient than the full projection. As ground state conical intersections are typically dominated by single excitations, doubly excited may vary more rapidly. Restricting the projection to the singles subspace leaves a better-conditioned effective Hamiltonian and a more stable solution, without affecting the amplitude equations or the properties discussed above. Details about all relevant terms can be found in previous publications \cite{SCCSD, SCC2, GCC}. The GCC equations are solved with the algorithm described in the original GCC paper \cite{GCC}. For the CVX-SCC equations, we apply the SCC DIIS solver used for the fixed-t SCC methods. The start guess for a single point calculation is set to the GCC full matrix eigenvectors and $\zeta = 0$. For successive calculations, it is possible to restart both the GCC optimization and the SCC DIIS solver using the parameters of a previous calculation as initial guesses. The DIIS results are checked with a Davidson solver, as for fixed-t SCC. The computational scaling of the CVX-SCC methods is the same as their standard CC counterparts, $N^5$ for CVX-SCC2 and $N^6$ for CVX-SCCSD.




\clearpage 

%
\bibliography{science_template} 
\bibliographystyle{sciencemag}

%
%
%
%
%
%


\section*{Acknowledgments}
We thank Eirik F. Kjønstad and Sara Angelico for their work on the SCCSD code which was the foundation for our implementation of the fixed-t SCC and CVX-SCC methods. We thank Sara Angelico for her assistance in the determination of the gh-planes used in our calculations.
\paragraph*{Funding:}
This work was supported by the European Research Council (ERC) under the European Union's Horizon 2020 Research and Innovation Program (grant agreement No. 101020016).
\paragraph*{Author contributions:}
L.S., F.R., and H.K. conceived the CVX-SCC framework and analyzed the data. L.S. and F.R. developed the implementation in eT and performed all the calculations. L.S. and F.R. wrote the first draft of the manuscript, and all authors reviewed and approved the final version. H.K. supervised the project and secured funding.
\paragraph*{Competing interests:}
There are no competing interests to declare.
\paragraph*{Data and materials availability:}
Calculation results are available at https://zenodo.org/records/22123560.


\subsection*{Supplementary materials}
Supplementary Text\\
Figures S1 to S2\\
Tables S1 to S25\\


\newpage


\renewcommand{\thefigure}{S\arabic{figure}}
\renewcommand{\thetable}{S\arabic{table}}
\renewcommand{\theequation}{S\arabic{equation}}
\renewcommand{\thepage}{S\arabic{page}}
\setcounter{figure}{0}
\setcounter{table}{0}
\setcounter{equation}{0}
\setcounter{page}{1} 


\begin{center}
\section*{Supplementary Materials for\\ \scititle}

	Leo Stoll$^{1\dagger}$,
	Federico Rossi$^{1\dagger}$,
	Henrik Koch$^{1\ast}$ \\
	\small$^{1}$Department of Chemistry, Norwegian University of Science and Technology, Trondheim \& 7050, Norway. \\
	\small$^\ast$Corresponding author. Email: henrik.koch@ntnu.no \\
	\small$^\dagger$These authors contributed equally to this work.
\end{center}

\subsection*{This PDF file includes:}
Supplementary Text\\
Figures S1 to S2\\
Tables S1 to S25\\


\newpage



\subsection*{Supplementary text}

\subsubsection*{Geometries for ethylene S$_0$/S$_1$ conical intersection}

The ethylene S$_0$/S$_1$ conical intersections were located in the gh-plane of the CCSD S$_0$/S$_1$ MECI, determined with the gradient projection method \cite{GCC, angelico2025determining}. The g- and h-vectors were calculated in units of Hartree/Bohr at the CCSD/cc-pVDZ level in a local branch of the electronic structure program eT \cite{angelico2025determining, kjonstad2023communication}, and were orthogonalized and rescaled for a suitable depiction of the intersection. The intersection geometries and g- and h-vectors applied in the 2D-scans are provided in Tables 
\ref{tab:sup_EthyleneS0S1_geoms_CCSD_CI}, 
\ref{tab:sup_EthyleneS0S1_geoms_CCSD_g}, 
\ref{tab:sup_EthyleneS0S1_geoms_CCSD_h}, 
\ref{tab:sup_EthyleneS0S1_geoms_CC2_CI},
\ref{tab:sup_EthyleneS0S1_geoms_CC2_g},
\ref{tab:sup_EthyleneS0S1_geoms_CC2_h}.

\subsubsection*{Uracil}

The uracil S$_1$/S$_2$ conical intersections where located in the gh-plane of the CCSD/cc-pVDZ S$_1$/S$_2$ MECI, determined with the gradient projection method \cite{angelico2025determining}. The g- and h-vectors were calculated in units of Hartree/Bohr at the CCSD/cc-pVDZ level in a local branch of the electronic structure program eT \cite{kjonstad2023communication}, and were orthogonalized and rescaled for a suitable depiction of the intersection. For the SCC2/CC2 scan, the g- and h-vectors were rotated by 15 degrees for a better rendition of the intersection. The intersection geometries and g- and h-vectors applied in the 2D-scans are provided in Tables 
\ref{tab:sup_UracilS1S2_geoms_CCSD_CI}, 
\ref{tab:sup_UracilS1S2_geoms_CCSD_g}, 
\ref{tab:sup_UracilS1S2_geoms_CCSD_h}.
\ref{tab:sup_UracilS1S2_geoms_CC2_CI}, 
\ref{tab:sup_UracilS1S2_geoms_CC2_g}, 
\ref{tab:sup_UracilS1S2_geoms_CC2_h}.

The uracil S$_0$/S$_1$ conical intersection was located in the gh-plane of the CCSD/cc-pVDZ $6S^5$ S$_0$/S$_1$ $\varepsilon$-MECI ($\varepsilon$ = 0.0025 eV) \cite{angelico2025determining}. The g- and h-vectors were calculated in units of Hartree/Bohr at the CCSD/cc-pVDZ level in a local branch of the electronic structure program eT \cite{angelico2025determining, kjonstad2023communication}, and were orthogonalized and rescaled for a suitable depiction of the intersection. The intersection geometry and g- and h-vectors applied in the 2D-scan are provided in Tables \ref{tab:sup_uracilS0S1_geoms_CCSD_CI}, \ref{tab:sup_UracilS0S1_geoms_CCSD_g}, \ref{tab:sup_UracilS0S1_geoms_CCSD_h}.

The non-adiabatic dynamics simulation was run in FMS90, with electronic structures from a development branch of the electronic structure program eT \cite{kjonstad2024photoinduced, angelico2026spectroscopic}. Electronic structures were obtained with the cc-pVDZ basis set.

\subsubsection*{Geometries for PSB3 S$_0$/S$_1$ calculations}
The PSB3 CVX-SCC2 S$_0$/S$_1$ conical intersection was determined in the gh-plane of the CC2 MECI geometry found in \cite{tuna2015assessment}. The g- and h-vectors were calculated in units of Hartree/Bohr at the CCSD/6-31G* level in a local branch of the electronic structure program eT \cite{angelico2025determining, kjonstad2023communication}, and were orthogonalized and rescaled for a suitable depiction of the intersection. The intersection geometry and g- and h-vectors are provided in Tables \ref{tab:sup_PSB3S0S1_geoms_CC2_CI}, \ref{tab:sup_PSB3S0S1_geoms_CC2_g}, \ref{tab:sup_PSB3S0S1_geoms_CC2_h}. 

\begin{figure}[htbp]
    \centering
    \includegraphics[width=\linewidth]{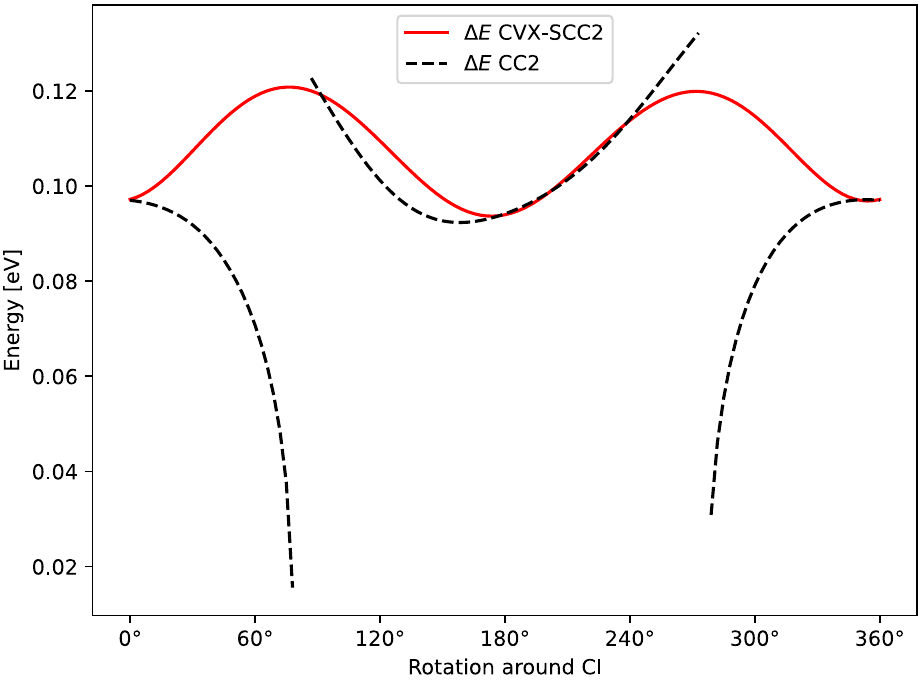}
    \caption{\textbf{S$_0$/S$_1$ energy differences of PSB3, when circumnavigating the S$_0$/S$_1$ conical intersection.} CVX-SCC2 results are shown in red and CC2 results are represented with black dashed line. All energies are expressed in eV. The circle is centered at the CVX-SCC2 CI with a radius of 0.02\AA.}
    \label{fig:sup_retinal_circle_diff}
\end{figure}

The provided intersection geometry was the center for the circular path used in Figure \ref{fig:PSB3_circle}, which circumnavigated the CI geometry at a distance of 0.02 \AA. A plot of the CVX-SCC2 and CC2 S$_1$ energy differences is shown in Figure \ref{fig:sup_retinal_circle_diff}.

\subsubsection*{HeH$_2$ circular path, intersection geometry and 2D-scan}

The approximate S$_0$/S$_1$ CI from the literature \cite{GPECC} was used as the center for the circular path in Figure \ref{fig:HeH2_circle}. The path moved around this center point at a distance of 0.05\AA in the CCSD/aug-cc-pVDZ branching plane \cite{GPECC}.

\begin{figure}[htbp] 
    \centering 
    \includegraphics[ width=\linewidth, trim=0 0 0 300, clip ]{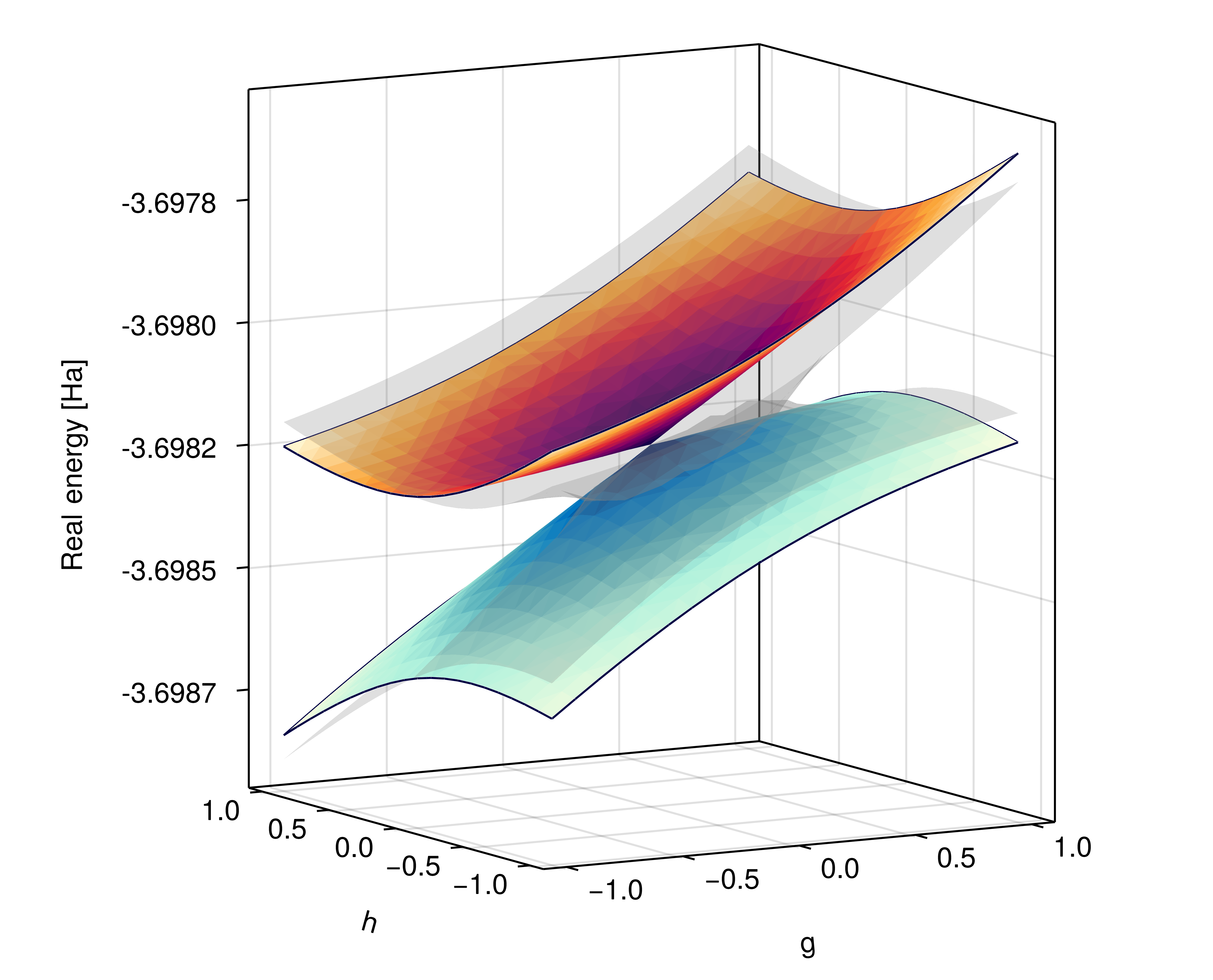} 
    \caption{\textbf{HeH$_2$ S$_0$/S$_1$ conical intersection of CVX-SCCSD/cc-pVDZ (colored) and GCCSD/cc-pVDZ (gray) with a CVX-HF reference state.}} \label{fig:sup_HeH2_CI} 
\end{figure}

A 2D-scan of the corresponding CVX-SCCSD/cc-pVDZ S$_0$/S$_1$ CI is shown in Figure \ref{fig:sup_HeH2_CI}. This intersection geometry was determined in the CCSD/aug-cc-pVDZ gh-plane from the literature \cite{GPECC}. The g- and h-vectors were orthogonalized and rescaled for a suitable depiction of the intersection. The intersection geometry and applied g- and h-vectors are provided in Tables \ref{tab:sup_HeH2_S0S1_geoms_CC2_CI}, \ref{tab:sup_HeH2_S0S1_geoms_CC2_g} and \ref{tab:sup_HeH2_S0S1_geoms_CC2_h}.

\subsubsection*{Vertical excitation calculations}

Results of vertical excitation calculations are shown in Table \ref{tab:sup_EnergyComparison}. The CCSD/cc-pVTZ equilibrium geometries are provided in Tables \ref{tab:sup_ethylene_equ}, \ref{tab:sup_Uracil_equ}, \ref{tab:sup_PSB3_equ}.


\begin{table} 
	\centering
	\caption{\textbf{Ground state and vertical excitation energies of the fixed-t SCC, GCC and CVX-SCC methods, compared with standard CC results.}
		Calculations were run at the CCSD/cc-pVTZ ground state equilibrium geometries, all using the cc-pVTZ basis set. The geometries are provided in the SI. The coupled cluster equations were converged to a residual threshold of $10^{-8}$ a.u. For fixed-t SCC, S$_1$ and S$_2$ were selected as the similarity constrained states. For GCC, one state was included in the GCC projection. For CVX-SCC, one state was included in the GCC projection, and the S$_0$ and S$_1$ states were selected as the similarity constrained states.}
	\label{tab:sup_EnergyComparison}
	
	\begin{tabular}{l r r r r r r}
		\\
        \hline
		System & Energy & CC2 & fixed-t SCC2 & GCC2 & CVX-SCC2 & CC3\\
		\hline
        Ethylene & $E_0$ [Ha] & -78.302449 & -78.302453 & -78.302449 & -78.302450 & -78.347412 \\
         & $\omega_1$ [eV] & 2.069573 & 2.069648 & 2.069697 & 2.069700 & 2.118620\\
         & $\omega_2$ [eV] & 6.361877 & 6.361917 & 6.361839 & 6.361841 & 5.795603\\
        \hline
        Uracil & $E_0$ [Ha] & -414.236808 & -414.236808 & -414.236808 & -414.236808 & -414.301771\\
         & $\omega_1$ [eV] & 5.032794 & 5.032798 & 5.032798 & 5.032798 & 5.041140\\
         & $\omega_2$ [eV] & 5.610163 & 5.610168 & 5.610168 & 5.610168 & 5.544615\\
		\hline
        PSB3 & $E_0$ [Ha] & -249.377468 & -249.377468 & -249.377469 & -249.377462  & -249.462257\\
         & $\omega_1$ [eV] & 4.256516 & 4.256519 & 4.260200 & 4.259860 & 4.261257\\
         & $\omega_2$ [eV] & 6.509952 & 6.509957 & 6.511152 & 6.510957 & 5.918578\\
		\hline
	\end{tabular}
	
	\begin{tabular}{l r r r r r r}
		\\
        \hline
		System & Energy & CCSD & fixed-t SCCSD & GCCSD & CVX-SCCSD & CC3\\
        \hline
        Ethylene & $E_0$ [Ha] & -78.333525 & -78.333476 & -78.333525 & -78.333525 & -78.347412\\
         & $\omega_1$ [eV] & 2.117914 & 2.115986 & 2.117851 & 2.117874 & 2.118620\\
         & $\omega_2$ [eV] & 6.334276 & 6.333073 & 6.334337 & 6.334346 & 5.795603\\
        \hline
        Uracil & $E_0$ [Ha] & -414.222217 & -414.222217 & -414.222217 & -414.222217 & -414.301771 \\
         & $\omega_1$ [eV] & 5.299818 & 5.299822 & 5.299823 & 5.299822 & 5.041140\\
         & $\omega_2$ [eV] & 5.803354 & 5.803358 & 5.803358 & 5.803358 & 5.544615\\
		\hline
        PSB3 & $E_0$ [Ha] & -249.408847 & -249.408847 & -249.408852 & -249.408934 & -249.462257\\
         & $\omega_1$ [eV] & 4.422465 & 4.422469 & 4.438358 & 4.443422 & 4.261257\\
         & $\omega_2$ [eV] & 6.617051 & 6.617056 & 6.621022 & 6.623238 & 5.918578\\
		\hline
	\end{tabular}
\end{table}

\begin{table}[h]
\centering
\caption{\textbf{Intersection geometry for SCCSD 2D-scan of uracil S$_1$/S$_2$ conical intersection.} Coordinates are provided in Angstrom.}
\label{tab:sup_UracilS1S2_geoms_CCSD_CI}
\begin{tabular}{c r r r}
\hline
Atom & x & y & z \\
\hline
C &  1.618074102967 &  0.406172691555 & -0.118759380089 \\
C &  1.436219519008 & -1.026285939005 & -0.232421583335 \\
C & -0.979727830384 & -0.709958147181 &  0.083890574664 \\
C &  0.487621488989 &  1.253065037565 & -0.173472466125 \\
N &  0.197512434133 & -1.489665095170 &  0.127262095417 \\
N & -0.773406827006 &  0.573032003421 & -0.386426043479 \\
O & -2.054174666636 & -1.199395116302 &  0.377583343280 \\
O &  0.425917100754 &  2.496464234167 &  0.010278097184 \\
H &  2.203088634351 & -1.757481621192 & -0.495538706783 \\
H &  0.028314815632 & -2.472311587501 &  0.343706583519 \\
H & -1.585368820818 &  1.177017134285 & -0.278115820477 \\
H &  2.588062025820 &  0.836730180171 &  0.140011559226 \\
\hline
\end{tabular} \\
\end{table}

\begin{table}[h]
\centering
\caption{\textbf{The applied g-vector for SCCSD 2D-scan of uracil S$_1$/S$_2$ conical intersection.} Coordinates are provided in Angstrom.}
\label{tab:sup_UracilS1S2_geoms_CCSD_g}
\begin{tabular}{c r r r}
\hline
Atom & x & y & z \\
\hline
C & -0.000564656643 &  0.001459267023 &  0.000525297701 \\
C & -0.000280166248 & -0.000880454629 & -0.001139282215 \\
C & -0.000645131451 & -0.000238008620 & -0.000143050800 \\
C &  0.001010021344 & -0.000942452486 &  0.000765623330 \\
N &  0.000489837630 &  0.000633663498 & -0.000127495519 \\
N & -0.000061523633 &  0.000037614031 & -0.000194246259 \\
O & -0.000027358147 & -0.000037706137 &  0.000210499431 \\
O & -0.000047145001 &  0.000133972474 & -0.000355709892 \\
H &  0.000064180728 & -0.000054990954 &  0.000412481334 \\
H &  0.000006020486 & -0.000070340576 &  0.000141983997 \\
H & -0.000091672905 &  0.000012665081 & -0.000014902771 \\
H &  0.000147593840 & -0.000053228705 & -0.000081198337 \\
\hline
\end{tabular} \\
\end{table}

\begin{table}[h]
\centering
\caption{\textbf{The applied h-vector for SCCSD 2D-scan of uracil S$_1$/S$_2$ conical intersection.} Coordinates are provided in Angstrom.}
\label{tab:sup_UracilS1S2_geoms_CCSD_h}
\begin{tabular}{c r r r}
\hline
Atom & x & y & z \\
\hline
C & -0.000011274053 & -0.000068042565 & -0.000010500471 \\
C &  0.000077304451 &  0.000191810639 &  0.000025345480 \\
C &  0.000098605603 & -0.000188616181 &  0.000040578655 \\
C &  0.000071142153 & -0.000378231238 & -0.000022958017 \\
N & -0.000186861253 & -0.000012127878 & -0.000005190183 \\
N & -0.000008725561 &  0.000127281705 & -0.000013870643 \\
O & -0.000017651008 &  0.000054025595 & -0.000014319576 \\
O & -0.000031857271 &  0.000234770836 &  0.000039268285 \\
H &  0.000009523710 &  0.000015730072 &  0.000015852422 \\
H &  0.000001795023 &  0.000025962902 & -0.000011097822 \\
H & -0.000009378560 & -0.000009951472 & -0.000000991651 \\
\hline
\end{tabular}
\end{table}

\begin{table}[h]
\centering
\caption{\textbf{Intersection geometry for SCC2 2D-scan of uracil S$_1$/S$_2$ conical intersection.} Coordinates are provided in Angstrom.}
\label{tab:sup_UracilS1S2_geoms_CC2_CI}
\begin{tabular}{c r r r}
\hline
Atom & x & y & z \\
\hline
C &  1.620169799300 &  0.405838304100 & -0.119611468700 \\
C &  1.432924467100 & -1.033967486200 & -0.230708026200 \\
C & -0.983163246600 & -0.699460265300 &  0.082150228900 \\
C &  0.481207239400 &  1.275357198800 & -0.174314173300 \\
N &  0.205977098400 & -1.490721138000 &  0.127873578000 \\
N & -0.772786446900 &  0.566275922800 & -0.385182596500 \\
O & -2.053178694400 & -1.202119617900 &  0.377770651100 \\
O &  0.427708718900 &  2.483830361800 &  0.009173517300 \\
H &  2.202419430400 & -1.758157487700 & -0.497467902000 \\
H &  0.028204892300 & -2.473481600400 &  0.343908178400 \\
H & -1.584633874600 &  1.177503728900 & -0.278024214000 \\
H &  2.587282535100 &  0.836486021900 &  0.142430299700 \\
\hline
\end{tabular}
\end{table}

\begin{table}[h]
\centering
\caption{\textbf{The applied g-vector for SCC2 2D-scan of uracil S$_1$/S$_2$ conical intersection.} Coordinates are provided in Angstrom.}
\label{tab:sup_UracilS1S2_geoms_CC2_g}
\begin{tabular}{c r r r}
\hline
Atom & x & y & z \\
\hline
C & -0.0014155380 &  0.0039935927 &  0.0013892993 \\
C & -0.0009884244 & -0.0029298003 & -0.0030220309 \\
C & -0.0020020111 &  0.0000378363 & -0.0005085043 \\
C &  0.0023561093 & -0.0011223252 &  0.0020513206 \\
N &  0.0019065682 &  0.0016740442 & -0.0003104923 \\
N & -0.0001283615 & -0.0003423528 & -0.0004524735 \\
O & -0.0000095569 & -0.0002835616 &  0.0005916206 \\
O & -0.0000114993 & -0.0004650889 & -0.0010517501 \\
H &  0.0001324513 & -0.0001959290 &  0.0010077649 \\
H &  0.0000093131 & -0.0002707793 &  0.0004040204 \\
H & -0.0002037666 &  0.0000669645 & -0.0000349645 \\
H &  0.0003547121 & -0.0001625895 & -0.0000638221 \\
\hline
\end{tabular}
\end{table}

\begin{table}[h]
\centering
\caption{\textbf{The applied h-vector for SCC2 2D-scan of uracil S$_1$/S$_2$ conical intersection.} Coordinates are provided in Angstrom.}
\label{tab:sup_UracilS1S2_geoms_CC2_h}
\begin{tabular}{c r r r}
\hline
Atom & x & y & z \\
\hline
C & -0.0000802374 &  0.0000196263 &  0.0000340975 \\
C &  0.0001203358 &  0.0002793983 & -0.0000689835 \\
C &  0.0001237025 & -0.0003890189 &  0.0000635822 \\
C &  0.0002420012 & -0.0008282565 &  0.0000349117 \\
N & -0.0003102765 &  0.0000421724 & -0.0000232260 \\
N & -0.0000232259 &  0.0002497835 & -0.0000469059 \\
O & -0.0000369315 &  0.0001004658 & -0.0000058708 \\
O & -0.0000664243 &  0.0004674123 &  0.0000390347 \\
H &  0.0000250429 &  0.0000246951 &  0.0000733277 \\
H &  0.0000040910 &  0.0000428743 & -0.0000067401 \\
H & -0.0000276087 & -0.0000179136 & -0.0000034586 \\
H &  0.0000295330 &  0.0000087549 & -0.0000897624 \\
\hline
\end{tabular}
\end{table}

\begin{table}[h]
\centering
\caption{\textbf{Intersection geometry for CVX-SCCSD 2D-scan of ethylene S$_0$/S$_1$ conical intersection.} Coordinates are provided in Angstrom.}
\label{tab:sup_EthyleneS0S1_geoms_CCSD_CI}
\begin{tabular}{c r r r}
\hline
Atom & x & y & z \\
\hline
C &  0.617196314500 &  0.131926743189 & -0.070850406433 \\
C & -0.771325109363 &  0.363828725359 & -0.252295957175 \\
H &  1.055810836328 & -0.494035542985 &  0.740629350476 \\
H &  1.354853336083 &  0.452693055292 & -0.819575212326 \\
H & -1.391978657030 &  0.455235939437 &  0.669776528666 \\
H & -0.864677155662 & -0.809638944361 & -0.268093570269 \\
\hline
\end{tabular}
\end{table}

\begin{table}[h]
\centering
\caption{\textbf{The applied g-vector for CVX-SCCSD 2D-scan of ethylene S$_0$/S$_1$ conical intersection.} Coordinates are provided in Angstrom.}
\label{tab:sup_EthyleneS0S1_geoms_CCSD_g}
\centering
\begin{tabular}{c r r r}
\hline
Atom & x & y & z \\
\hline
C &  0.000047798054 &  0.000028186743 &  0.000078769870 \\
C & -0.000042364972 & -0.000113179958 &  0.000079340711 \\
H &  0.000019350600 &  0.000151205148 &  0.000098516626 \\
H & -0.000029225873 & -0.000117070522 & -0.000067842787 \\
H &  0.000079938375 & -0.000006616332 &  0.000080690594 \\
H & -0.000075496183 &  0.000057474921 & -0.000269475015 \\
\hline
\end{tabular}
\end{table}

\begin{table}[h]
\centering
\caption{\textbf{The applied h-vector for CVX-SCCSD 2D-scan of ethylene S$_0$/S$_1$ conical intersection.} Coordinates are provided in Angstrom.}
\label{tab:sup_EthyleneS0S1_geoms_CCSD_h}
\begin{tabular}{c r r r}
\hline
Atom & x & y & z \\
\hline
C & -0.000156447034 & -0.000083097672 &  0.000030951689 \\
C &  0.000035864886 &  0.000185187204 & -0.000098024110 \\
H &  0.000026415812 &  0.000011324035 &  0.000016101612 \\
H & -0.000000191196 & -0.000011501611 & -0.000014030270 \\
H &  0.000023363949 & -0.000072078907 &  0.000037637096 \\
H &  0.000071762896 & -0.000030208607 &  0.000027712856 \\
\hline
\end{tabular}
\end{table}

\begin{table}[h]
\centering
\caption{Intersection geometry for CVX-SCC2 2D-scan of ethylene S$_0$/S$_1$ conical intersection. Coordinates are provided in Angstrom.}
\label{tab:sup_EthyleneS0S1_geoms_CC2_CI}
\begin{tabular}{c r r r}
\hline
Atom & x & y & z \\
\hline
C &  0.620087782239 &  0.133472352828 & -0.071113783837 \\
C & -0.772097851517 &  0.360208050773 & -0.250311460312 \\
H &  1.055418554696 & -0.493703713880 &  0.740693647261 \\
H &  1.354754413823 &  0.452483744034 & -0.819568260199 \\
H & -1.392105819649 &  0.456467913118 &  0.669403457129 \\
H & -0.866190910917 & -0.808911831255 & -0.269518942093 \\
\hline
\end{tabular}
\end{table}

\begin{table}[h]
\centering
\caption{The applied g-vector for CVX-SCC2 2D-scan of ethylene S$_0$/S$_1$ conical intersection. Coordinates are provided in Angstrom.}
\label{tab:sup_EthyleneS0S1_geoms_CC2_g}
\begin{tabular}{c r r r}
\hline
Atom & x & y & z \\
\hline
C &  0.000039831712 &  0.000023488953 &  0.000065641559 \\
C & -0.000035304144 & -0.000094316632 &  0.000066117259 \\
H &  0.000016125500 &  0.000126004290 &  0.000082097188 \\
H & -0.000024354894 & -0.000097558768 & -0.000056535656 \\
H &  0.000066615312 & -0.000005513610 &  0.000067242162 \\
H & -0.000062913486 &  0.000047895767 & -0.000224562512 \\
\hline
\end{tabular}
\end{table}

\begin{table}[h]
\centering
\caption{The applied h-vector for CVX-SCC2 2D-scan of ethylene S$_0$/S$_1$ conical intersection. Coordinates are provided in Angstrom.}
\label{tab:sup_EthyleneS0S1_geoms_CC2_h}
\begin{tabular}{c r r r}
\hline
Atom & x & y & z \\
\hline
C & -0.000156447034 & -0.000083097672 &  0.000030951689 \\
C &  0.000035864886 &  0.000185187204 & -0.000098024110 \\
H &  0.000026415812 &  0.000011324035 &  0.000016101612 \\
H & -0.000000191196 & -0.000011501611 & -0.000014030270 \\
H &  0.000023363949 & -0.000072078907 &  0.000037637096 \\
H &  0.000071762896 & -0.000030208607 &  0.000027712856 \\
\hline
\end{tabular}
\end{table}

\begin{table}[h]
\centering
\caption{\textbf{Intersection geometry for CVX-SCCSD 2D-scan of uracil S$_0$/S$_1$ conical intersection.} Coordinates are provided in Angstrom.}
\label{tab:sup_uracilS0S1_geoms_CCSD_CI}
\begin{tabular}{c r r r}
\hline
Atom & x & y & z \\
\hline
N  & -0.677734400662 & -0.221446299279 & -0.060234581819\\
N  &  0.716095249916 &  1.667932663532 &  0.209280136114\\
C  & -0.604057174865 &  1.120002285951 &  0.275320515086\\
C  &  1.737778775810 &  0.872885104403 & -0.194945358790\\
C  &  1.695451271560 & -0.509329759550 &  0.265823839868\\
C  &  0.391008249594 & -1.180287567604 & -0.022709458403\\
O  &  0.128118841661 & -2.355050664423 & -0.103639453343\\
O  & -1.551773503379 &  1.835950954976 &  0.503385450271\\
H  & -1.622848700504 & -0.598517542855 & -0.092042647017\\
H  &  0.724591983831 &  2.681807589696 &  0.094448644995\\
H  &  1.757076876094 & -0.500660664956 &  1.375675736470\\
H  &  2.514411902730 &  1.327562972312 & -0.820013541020\\
\hline
\end{tabular} \\
\end{table}

\begin{table}[h]
\centering
\caption{\textbf{The applied g-vector for CVX-SCCSD 2D-scan of uracil S$_0$/S$_1$ conical intersection.} Coordinates are provided in Angstrom.}
\label{tab:sup_UracilS0S1_geoms_CCSD_g}
\begin{tabular}{c r r r}
\hline
Atom & x & y & z \\
\hline
N  &  0.000198601558 & -0.000183831041 &  0.000032448635 \\
N  & -0.000500994762 &  0.000003508440 &  0.000461237366 \\
C  &  0.000344980812 &  0.000422464947 & -0.000114220030 \\
C  &  0.000647774783 &  0.000088691590 & -0.000632475571 \\
C  & -0.000225215691 &  0.000258999624 &  0.000401600473 \\
C  & -0.000227322340 & -0.000300772233 & -0.000221887715 \\
O  &  0.000039290096 &  0.000002168758 & -0.000013123930 \\
O  & -0.000100978300 & -0.000066676339 &  0.000022933688 \\
H  &  0.000033369281 & -0.000036441800 &  0.000023510757 \\
H  &  0.000030694202 & -0.000063090557 &  0.000100223579 \\
H  & -0.000113765544 & -0.000121962054 & -0.000137332018 \\
H  & -0.000126434095 & -0.000003059336 &  0.000077084764 \\
\hline
\end{tabular} \\
\end{table}

\begin{table}[h]
\centering
\caption{\textbf{The applied h-vector for CVX-SCCSD 2D-scan of uracil S$_0$/S$_1$ conical intersection.} Coordinates are provided in Angstrom.}
\label{tab:sup_UracilS0S1_geoms_CCSD_h}
\begin{tabular}{c r r r}
\hline
Atom & x & y & z \\
\hline
N  & -0.000056403391 & -0.000054244748 & -0.000044759229 \\
N  & -0.001734903978 &  0.000049759214 & -0.000980556333 \\
C  &  0.000250094851 &  0.000425496867 & -0.000149741004 \\
C  &  0.000700838080 & -0.000560867284 &  0.000454565704 \\
C  & -0.000431164017 & -0.000083161227 & -0.000328083260 \\
C  & -0.000945802613 &  0.001250015973 &  0.000781702190 \\
O  &  0.000367546570 & -0.000104735044 & -0.000170074331 \\
O  & -0.000270426873 &  0.000017778125 &  0.000145809136 \\
H  & -0.000059028059 &  0.000007550174 &  0.000032861936 \\
H  & -0.000038227914 & -0.000171709616 & -0.000406345771 \\
H  &  0.000527624697 &  0.000167140778 &  0.000918772053 \\
H  &  0.001690204715 & -0.000943901807 & -0.000254830622 \\
\hline
\end{tabular}
\end{table}

\begin{table}[h]
\centering
\caption{\textbf{Intersection geometry for CVX-SCC2 2D-scan of PSB3 S$_0$/S$_1$ conical intersection.} Coordinates are provided in Angstrom.}
\label{tab:sup_PSB3S0S1_geoms_CC2_CI}
\begin{tabular}{c r r r}
\hline
Atom & x & y & z \\
\hline
C &   -2.869546755092 &  -0.892685385358 &  0.823938242784 \\
C &   -1.485675507739 &  -0.868679251171 &  0.818442369778 \\
C &   -0.723066410073 &  -0.014135806613 &  0.025205823354 \\
C &    0.739717991547 &  -0.057072862605 &  0.013985232756 \\
C &    1.516428936296 &   0.803018650220 &  0.875043733867 \\
N &    2.819869352697 &   0.774651071543 &  0.859183109864 \\
H &    1.320932055718 &  -0.720714137664 & -0.639565809707 \\
H &   -1.176453955112 &   0.653072119125 & -0.709349402746 \\
H &    1.015595126817 &   1.490865766602 &  1.553733102162 \\
H &   -0.967128403865 &  -1.557272312834 &  1.491401141955 \\
H &    3.375823548073 &   1.383951729799 &  1.469597788764 \\
H &    3.344891980574 &   0.144848396631 &  0.240451647527 \\
H &   -3.421889463341 &  -1.567251760716 &  1.466770599817 \\
H &   -3.434028677521 &  -0.230734489379 &  0.170851528854 \\
\hline
\end{tabular}
\end{table}

\begin{table}[h]
\centering
\caption{\textbf{The applied g-vector for CVX-SCC2 2D-scan of PSB3 S$_0$/S$_1$ conical intersection.} Coordinates are provided in Angstrom.}
\label{tab:sup_PSB3S0S1_geoms_CC2_g}
\centering
\begin{tabular}{c r r r}
\hline
Atom & x & y & z \\
\hline
C &  -0.000006808235 & -0.000027388581 &  0.000027092758 \\
C &  -0.000018977691 &  0.000008341446 & -0.000031832914 \\
C &  -0.000016794535 &  0.000043316611 &  0.000030681715 \\
C &  -0.000119374176 & -0.000133747773 & -0.000148568048 \\
C &   0.000278203596 &  0.000142064024 &  0.000134034977 \\
N &  -0.000212867422 & -0.000017115927 & -0.000019185195 \\
H &   0.000034484171 & -0.000002018973 &  0.000008731263 \\
H &   0.000033546036 & -0.000003260495 & -0.000007729002 \\
H &  -0.000005854126 & -0.000001140197 &  0.000000008887 \\
H &  -0.000007021661 & -0.000008911279 &  0.000006659671 \\
H &   0.000016517867 &  0.000020016558 &  0.000021458431 \\
H &   0.000014297333 & -0.000021431738 & -0.000020466856 \\
H &   0.000004999886 &  0.000005978662 & -0.000003158999 \\
H &   0.000005648957 & -0.000004702336 &  0.000002273312 \\
\hline
\end{tabular}
\end{table}

\begin{table}[h]
\centering
\caption{\textbf{The applied h-vector for CVX-SCC2 2D-scan of PSB3 S$_0$/S$_1$ conical intersection.} Coordinates are provided in Angstrom.}
\label{tab:sup_PSB3S0S1_geoms_CC2_h}
\begin{tabular}{c r r r}
\hline
Atom & x & y & z \\
\hline
C  &  0.000000419457 & -0.000004706922 & -0.000006759910 \\
C  &  0.000002603869 & -0.000161501264 & -0.000173354268 \\
C  &  0.000000689895 & -0.000024686648 &  0.000058767582 \\
C  &  0.000004180562 &  0.000021442620 & -0.000024898606 \\
C  & -0.000007983505 &  0.000189624424 & -0.000205707220 \\
N  &  0.000009010264 & -0.000027376100 &  0.000029272076 \\
H  & -0.000003677872 & -0.000146561776 &  0.000147035881 \\
H  & -0.000003781797 &  0.000150293746 &  0.000135231167 \\
H  &  0.000000169037 & -0.000003291821 &  0.000003267056 \\
H  &  0.000001111409 &  0.000000465689 & -0.000000394412 \\
H  & -0.000000884430 & -0.000019377941 &  0.000017884688 \\
H  & -0.000000594271 &  0.000003498832 & -0.000001777554 \\
H  & -0.000000823578 &  0.000017919569 &  0.000018602546 \\
H  & -0.000000365288 &  0.000003515612 &  0.000003212458 \\
\hline
\end{tabular}
\end{table}

\begin{table}[h]
\centering
\caption{\textbf{Intersection geometry for CVX-SCCSD 2D-scan of HeH$_2$ S$_0$/S$_1$ conical intersection.} Coordinates are provided in Angstrom.}
\label{tab:sup_HeH2_S0S1_geoms_CC2_CI}
\begin{tabular}{c r r r}
\hline
Atom & x & y & z \\
\hline
He & -3.077370227526 & 2.124946347247 & 0.000000000000 \\
H  & -3.092902858160 & 2.823795331019 & 0.000000000000 \\
H  & -1.369332820636 & 1.714975586462 & 0.000000000000 \\
\hline
\end{tabular}
\end{table}

\begin{table}[h]
\centering
\caption{\textbf{The applied g-vector for CVX-SCCSD 2D-scan of HeH$_2$ S$_0$/S$_1$ conical intersection.} Coordinates are provided in Angstrom.}
\label{tab:sup_HeH2_S0S1_geoms_CC2_g}
\centering
\begin{tabular}{c r r r}
\hline
Atom & x & y & z \\
\hline
He & -0.000046324065 &  0.000400980666 &  0.000000000000 \\
H  &  0.000063097469 & -0.000428308810 &  0.000000000000 \\
H  & -0.000016773403 &  0.000027328143 &  0.000000000000 \\
\hline
\end{tabular}
\end{table}

\begin{table}[h]
\centering
\caption{\textbf{The applied h-vector for CVX-SCCSD 2D-scan of HeH$_2$ S$_0$/S$_1$ conical intersection.} Coordinates are provided in Angstrom.}
\label{tab:sup_HeH2_S0S1_geoms_CC2_h}
\begin{tabular}{c r r r}
\hline
Atom & x & y & z \\
\hline
He & -0.001033496628 & -0.000278178583 & 0.000000000000 \\
H  &  0.000804968394 & -0.000019996837 & 0.000000000000 \\
H  &  0.000228438987 &  0.000298003343 & 0.000000000000 \\
\hline
\end{tabular}
\end{table}

\begin{table}[h]
\centering
\caption{\textbf{CCSD/cc-pVTZ ground state equilibrium geometry of ethylene.} Coordinates are provided in Angstrom.}
\label{tab:sup_ethylene_equ}
\begin{tabular}{c r r r}
\hline
Atom & x & y & z \\
\hline
C   &   0.375237812106  &  -0.066746315119  &  -0.128623175387 \\
C   &  -0.824773846133  &   0.768756971767  &  -0.110233797140 \\
H   &   0.821507087708  &  -0.380752817493  &   0.813459607724 \\
H   &   1.119720005768  &   0.214582613298  &  -0.864119197496 \\
H   &  -1.346795998088  &   0.522630585910  &   0.826523240306 \\
H   &  -0.145028892695  &  -0.958454522742  &  -0.537422020044 \\
\hline
\end{tabular}
\end{table}

\begin{table}[h]
\centering
\caption{\textbf{CCSD/cc-pVTZ ground state equilibrium geometry of uracil.} Coordinates are provided in Angstrom.}
\label{tab:sup_Uracil_equ}
\begin{tabular}{c r r r}
\hline
Atom & x & y & z \\
C   &   1.609194975446  &   0.367785454719  &  -0.145863909131 \\
C   &   1.416623711069  &  -0.951892316271  &  -0.065528562708 \\
C   &  -0.989039349669  &  -0.749889376292  &   0.047803630497 \\
C   &   0.462747580330  &   1.262654077118  &  -0.134458375763 \\
N   &   0.164523281665  &  -1.500195680914  &   0.028384934522 \\
N   &  -0.760141146650  &   0.601880263208  &  -0.036108587271 \\
O   &  -2.087066119479  &  -1.238098300793  &   0.130449710143 \\
O   &   0.506782056133  &   2.466152758504  &  -0.200931916323 \\
H   &   2.228323558400  &  -1.656035311611  &  -0.069899203899 \\
H   &   0.033642866385  &  -2.490766656980  &   0.088058165067 \\
H   &  -1.581218807696  &   1.181413686387  &  -0.024675414298 \\
H   &   2.587759370912  &   0.794375177681  &  -0.219232217761 \\
\hline
\end{tabular}
\end{table}

\begin{table}[h]
\centering
\caption{\textbf{CCSD/cc-pVTZ ground state equilibrium geometry of PSB3.} Coordinates are provided in Angstrom.}
\label{tab:sup_PSB3_equ}
\begin{tabular}{c r r r}
\hline
Atom & x & y & z \\
C  &   -2.714213277505   & -0.809415557392  &   0.867214793394 \\
C  &   -1.432278231655   & -0.454680259541  &   1.005454052720 \\
C  &   -0.622636167537   & -0.207725154296  &  -0.155382440353 \\
C  &    0.683140802871   &  0.157237917716  &  -0.211317897655 \\
C  &    1.462211263215   &  0.353152965534  &   0.945640360051 \\
N  &    2.709660183915   &  0.701261779368  &   0.922630231958 \\
H  &    1.152271117946   &  0.305549467166  &  -1.167823400885 \\
H  &   -1.128515786893   & -0.331367148157  &  -1.100334321729 \\
H  &    1.032036363175   &  0.215316822227  &   1.922921444547 \\
H  &   -1.013314620952   & -0.355993167566  &   1.990846314700 \\
H  &    3.232416672706   &  0.831377346658  &   1.772283102946 \\
H  &    3.196565561587   &  0.852922402842  &   0.053507857614 \\
H  &   -3.339947014600   & -0.999479110062  &   1.721818847216 \\
H  &   -3.161927047345   & -0.916296576905  &  -0.107769835547 \\
\hline
\end{tabular}
\end{table}

\clearpage



\end{document}